\documentstyle{mn}

\input{epsf}
\begin{document}

\title[Cosmological Perturbation Theory and the Spherical Collapse model]
{Cosmological Perturbation Theory and the Spherical Collapse model-
I. Gaussian initial conditions.} 

\author[P.Fosalba and E.Gazta\~{n}aga]
{Pablo Fosalba and Enrique Gazta\~{n}aga \\ 
Institut d'Estudis Espacials de Catalunya, Research Unit (CSIC), \\
Edf. Nexus-201 - c/ Gran Capit\`a 2-4, 08034 Barcelona, Spain}

\maketitle
 
\def\Mpc{{\,h^{-1}\,{\rm Mpc}}}
\def\mpc {h^{-1} {\rm{Mpc}}}
\def\and  {\it {et al.} \rm}
\def\rmd {\rm d}

\def\etal{{\it et al. }}
\def\ie {{\it i.e., }} 
\def\eg {{\it e.g., }}
\def\spose#1{\hbox to 0pt{#1\hss}}
\def\simlt{\mathrel{\spose{\lower 3pt\hbox{$\mathchar"218$}}
     \raise 2.0pt\hbox{$\mathchar"13C$}}}
\def\simgt{\mathrel{\spose{\lower 3pt\hbox{$\mathchar"218$}}
     \raise 2.0pt\hbox{$\mathchar"13E$}}}
\def\beq{\begin{equation}}
\def\eeq{\end{equation}}
\def\bce{\begin{center}}
\def\ece{\end{center}}
\def\bea{\begin{eqnarray}}
\def\eea{\end{eqnarray}}
\def\ben{\begin{enumerate}}
\def\een{\end{enumerate}}
\def\ul{\underline}
\def\ni{\noindent}
\def\nn{\nonumber}
\def\bs{\bigskip}
\def\ms{\medskip}
\def\wt{\widetilde}
\def\brr{\begin{array}}
\def\err{\end{array}}
\def\dsp{\displaystyle}
\newcommand{\rhobar}{\overline{\rho}}
\newcommand{\rhohat}{\hat{\rho}}
\newcommand{\xibar}{\overline{\xi}}
\newcommand{\deltabar}{\overline{\delta}}
\newcommand{\sigmabar}{\overline{\sigma}}
\newcommand{\deltahat}{\hat{\delta}}
\newcommand{\sigmahat}{\hat{\sigma}}
\def\Or{{\cal O}}
\def\calG{{\cal G}}
\def\La{{\cal L}}
\def\dotD {{\dot{D}}} 
\def\dotF {{\dot{F}}} 
\def\ddotF {{\ddot{F}}}
\def\ddotR {{\ddot{R}}}
\def\dotdel {{\ddot{\delta}}}
\def\ddotdel {{\ddot{\delta}}} 
\def\xdot {{\dot{\bf x }} }
\def\dota {{\dot{a}}} 
\def\vpsi {{\bf \Psi }}

\font\twelveBF=cmmib10 scaled 1200
\newcommand{\bte}{\hbox{\twelveBF $\theta$}}
\newcommand{\x}{\hbox{\twelveBF x}}
\newcommand{\q}{\hbox{\twelveBF q}}
\newcommand{\vv}{\hbox{\twelveBF v}}
\newcommand{\y}{\hbox{\twelveBF y}}
\newcommand{\r}{\hbox{\twelveBF r}}
\newcommand{\k}{\hbox{\twelveBF k}}
\newcommand{\lexp}{\mathop{\bigl\langle}}
\newcommand{\rexp}{\mathop{\bigr\rangle}}
\newcommand{\rexpc}{\mathop{\bigr\rangle_c}{}}
\newcommand{\eq}{{equation~}}

\begin{abstract}
We present a simple and intuitive approximation
for solving perturbation theory (PT) of small cosmic
fluctuations. We consider only the spherically symmetric or
monopole contribution to the PT integrals, which yields the exact result
for tree-graphs (e.g. at leading order).
We find that the non-linear evolution in Lagrangian space
is then given by a simple {\it local} transformation over the initial
conditions, although it is not local in Euler space. 
This transformation is found to be described 
the spherical collapse (SC) dynamics, as it is the exact solution 
in the shearless (and therefore local) approximation
in Lagrangian space. 
Taking advantage of this property, it is straightforward 
to derive the one-point 
cumulants, $\xi_J$, for both the unsmoothed and
smoothed density fields to {\it arbitrary} order
in the perturbative regime. To leading order 
this reproduces, and provides with a simple explanation for, the
exact results obtained by Bernardeau (1992, 1994).
We then show that the SC model
leads to accurate estimates for the next
corrective terms when compared to the results
derived in the exact perturbation theory making use of
the loop calculations (Scoccimarro \& Frieman 1996).
The agreement is within a few per cent
for the hierarchical ratios $S_J= \xi_J/\xi_2^{J-1}$.
We compare our analytic results to N-body simulations, which turn 
out to be in very good agreement up to scales where
$\sigma \approx 1$.  
A similar treatment is presented to estimate higher order corrections
in the Zel'dovich approximation. 
These results represent a powerful and readily usable tool
to produce analytical predictions to describe the gravitational
clustering of large scale structure in the weakly non-linear regime.  

\end{abstract}







\section{Introduction}

One of the important problems in Cosmology
is the origin of galaxies, clusters, and structures seen on huge
scales in the spatial distribution of galaxies and
in the microwave background radiation. 
Given some initial seed fluctuations we would like to be able to make
analytic predictions to describe the gravitational
clustering that generates the
large-scale structure that we see today.

The probability distribution function contains all the statistical 
information concerning the cosmic fields: density 
$\delta$ and velocity $v$ fluctuations. Here we will concentrate
on its moments, the variance, skewness, kurtosis and so on.
For a given set of moments in the initial conditions (IC), we would like
to derive the final moments using
the exact dynamics that rule the evolution of the underlying field.
The problem is that the
exact solution to the dynamical equations is only known for 
the linear regime of the evolution of these fields and only approximate 
solutions are known for their corresponding non-linear stages.  
The spherical infall model has been used to provide a qualitative 
picture of how isolated (gravitationally bound) structures evolve
up to their final collapse. In particular, the spherical model
gives an estimation of when the first astrophysical objects formed
purely under influence of gravity (see Pebbles 1980, and references
therein). 
This motivated the
development of the so-called Press-Schechter formalism 
(Press \& Schechter 1974) which gives the distribution of 
collapsed objects as a function of their mass.  
It is commonly assumed that highly non-linear objects such as galaxies
form at peaks (\ie values above a threshold)
of the underlying matter density field, what provides a possible mechanism 
for the segregation ({\em bias}) between luminous and dark matter
in the universe (Kaiser 1984, Bardeen \etal 1986).    
More recently, Mo \& White (1996) have extended the spherical collapse model
to build a model for the spatial clustering of non-linear
objects such as dark matter halos. 
Here we shall explain how the spherical collapse model also
appears in the context of Cosmological Perturbation Theory. 

Cosmological Perturbation theory (PT) provides a 
framework to study small departures
from linear theory: the weakly non-linear regime. 
The leading order 
contribution to the 
skewness for Gaussian initial conditions (GIC)
was obtained by Peebles (1980). Fry (1984) extended this
result to higher order moments and
Bernardeau (1992, hereafter B92) 
found the generating function to the
leading order contribution: the tree-level solution.

Comparison with observations and simulations made it necessary
to develop PT for the smoothed fields. 
The smoothing corrections to 
the unsmoothed amplitudes for a power-law
were first computed for the skewness for 
either a top-hat or a Gaussian window function by Juskiewicz \etal (1993).
For a top-hat filter, Bernardeau 1994a  (hereafter B94a)
developed the machinery to systematically derive the smoothed
hierarchical amplitudes for either the density ($S_J$) or the velocity 
divergence ($T_J$) fields to an arbitrary order (and arbitrary
power spectrum). These results are in excellent agreement
with numerical simulations (\eg Baugh, Gazta\~{n}aga,  Efstathiou 1995, 
hereafter BGE95; Gazta\~{n}aga \& Baugh 1995, Colombi \etal 1996).
Work by Matsubara (1994) and $\L$okas \etal (1995) include some results
for the Gaussian-smoothed density and velocity cumulants.

The success of the PT approach made it plausible to
go further and study higher order (loop) corrections and the case
of  non-Gaussian initial conditions (NGIC). 
Despite the impressive achievements in the diagrammatic approach
to loop corrections in PT by Scoccimarro $\&$ Frieman 1996a and 1996b, 
(hereafter
 SF96a and  SF96b, respectively), the
analytic entanglement faced when carrying out the calculations
for the smoothed one-point cumulants is enormous. 
Moreover, regularization
techniques must be used to evaluate the loop corrections through the 
kernels in Fourier space which, for a value of the spectral index 
$n = -1$ 
lead to logarithmic divergences which become powers of the cut-off used 
in the regularization for $n > -1$. This puts constraints on 
the domain of applicability of the exact PT itself. 
When it comes to calculating the statistics of the smoothed 
fields, further analytic entanglement makes comparison with simulations 
and observations be restricted to few discrete values of the spectral index
(Scoccimarro 1997, hereafter S97). 
For non-Gaussian initial conditions we have to face similar problems 
to start with, as loop corrections could enter at 
the same order than the tree-level contribution.

The general results derived
by Bernardeau (B92, B94a and Bernardeau 1994b, hereafter B94b)
show that the generating function follows the equation 
for the spherical collapse model. In this context 
this remarkable result lacked a satisfactory interpretation.
What is the connection between PT and the spherical collapse
model? Why does the generating function follow an equation
for a fluctuating field? Why is the smoothed generating function
given by the same unsmoothed result but
with a different argument? Is this
related to the Gaussian nature of the IC or the flatness of space?
In this paper we will present a
simple interpretation for these results, which will allow
us to extend it to higher-order corrections and
non-Gaussian initial conditions (NGIC).

This paper is organized as follows.
We present the monopole approximation to PT in \S\ref{sec:mon_pt}.
In section \S\ref{sec:PT} we give 
a brief account of the estimation of the cumulants in PT 
and some of its nonlinear approximations. 
We investigate the Spherical Collapse (SC)
model and its relation to PT in \S\ref{sec:sc}. 
In section \S\ref{sec:comparison}
we present a comparison of the SC model with exact PT and N-body Simulations.
General formulas for the one-point cumulants from a local transformation
of the density field in {\em Lagrangian} space (for GIC) are provided
in Appendix \ref{app:locumul}. 
A comparison with analog results in {\em Euler} space is given 
in Appendix \ref{app:euler}. 
An alternative derivation of the smoothing effects 
in Fourier space for a top-hat window is illustrated in Appendix 
\ref{app:nu2}. 
Further results are derived for the Spherically Symmetric Zel'dovich 
Approximation in Appendix \ref{app:za}.

\section{Perturbation Theory and the Monopole Approximation}
\label{sec:mon_pt}

We will now review briefly the main features of perturbation 
theory (PT) to present
the spherical symmetric solution or monopole approximation.
Throughout this paper we shall focus on Gaussian initial conditions and
the Einstein-deSitter scenario (FRW with $\Omega = 1$, $\Lambda = 0$) 
to gain simplicity. However, a
generalisation of the current approach to non-Gaussian IC and non-flat 
FRW universes will be 
provided in two accompanying papers (see Gazta\~{n}aga \& Fosalba 1998, 
Fosalba \& Gazta\~{n}aga 1998,
Papers  II \& III, hereafter). 

\subsection{Equations of Motion}

In a non-relativistic presureless (collisionless) expanding universe, 
the field equations for the local fluctuation 
field $\delta(x,t)$ in Euler space 
are:
 
\bea
{\partial\delta\over{\partial t}}+ {1\over{a}}\,\nabla \cdot
\left[(1+\delta)\, \vv\right] &=& 0 \label{continu}
 \\ \nonumber
{\partial\vv\over{\partial t}}+ {\dot{a}\over{a}}\,\vv +
{1\over{a}}(\vv\cdot\nabla)\cdot \vv &=&  - {1\over{a}} \, \nabla \phi \\
\nabla^2 \phi &=& 4 \pi \, G \, \bar{\rho} \, a^2 \, \delta
\label{fieldeq}
\eea
which correspond to the mass conservation, Euler and Poisson equations
respectively. The expansion factor $a=a(t) \equiv (1+z)^{-1}$  obeys:
\bea
{{\dot{a}}\over{a}} &\equiv& H ~=~ H_0 \left[ \Omega_0/a^3 + 
(1-\Omega_0-\Omega_{\Lambda})/a^2 + \Omega_{\Lambda} \right]^{1/2} 
\nonumber \\
\Omega_0 &=& {{8 \pi G \rho_0}\over{3 H_0^2}} ~;~
\Omega_{\Lambda} = {{\Lambda}\over{3 H_0^2}}.
\label{cosmo}
\eea
Here  $H_0$ and $\rho_0$ are the present values 
of the Hubble constant and
background density, and $\Lambda$ is the 
cosmological constant (see \eg Peebles 1980).
 
\subsection{Perturbation Theory Solutions}
\label{subsec:PT}

A convenient approach to solve the field equations above
is to expand the density 
contrast $\delta$ assuming it is small as is usually done in the 
context of perturbation theory (PT) in Euler space,
\beq
\delta(\x,t) = \delta_1(\x,t) 
+ \delta_2(\x,t)+ \delta_3(\x,t) ..., 
\label{pt}
\eeq
were we assume that $\delta_n << \delta_{n-1}$.
The first term $\delta_1 \equiv \delta_l$
is the solution to the linearized field equations and is
given in terms of the linear growth factor
$D(t)$. The second term
$\delta_2 \sim \delta_l^2$ is the 
second-order solution, obtained by using the linear solution in the 
source terms, and so on.

The dominant mode to the
linear  PT solution, $\delta_l$, 
 is given in terms of the linear growth factor
$D(t)$:
\bea
\delta_l(x,t) &=& D(t) \,  ~\delta(x,0) \nonumber \\ 
D(t) &=& {{\dota\over{a}}} ~\int_0^a da~\dota^{-3},
\label{Dt}
\eea
which for $\Omega=1$, the so-called Einstein-deSitter universe, gives 
$D(t)=a(t)$.  The linear approximation is therefore trivially
local: {\it the evolution of a density fluctuation at a given point is not affected 
by its 
neighboring density fluctuations},  
and all fluctuations grow with time
at the same rate.

Beyond linear order, 
the equations of motion can be integrated for an Einstein-de Sitter
universe, yielding for the growing mode (Peebles 1980),
\beq
\delta_2 = {5\over 7}\,\delta_l^2\,+\,\delta_{l,\alpha}
\,\Delta_{,\alpha}\,+\,{2\over 7}\,\Delta_{,\alpha \beta}\,
\Delta_{,\alpha \beta}
\label{secpert}
\eeq
where $_{,\alpha}$ are partial space derivatives and we denote,
$\Delta (\x) = -\,4\pi\,\int {d^{3}\x^\prime \,\delta (\x^\prime) /
|\x\,-\,\x \prime|}$.


The second order solution Eq.[\ref{secpert}],
explicitly shows that, although $\delta_2$ is of order
$\delta_l^2$, PT is formally {\em non-local} already
at the second order as explicitly realized by the term accounting for the
potential generated
by the density fluctuations at different points ($\sim \Delta$). In fact, the 
non-local
nature of gravity is an intrinsic feature of General Relativity
and remains so in the Newtonian limit through the tidal forces when the  
appropriate limit to the equations of motion is taken as was recently shown
by Kofman \& Pogosyan (1995).

To win simplicity in the equations of motion it is convenient to
work in Fourier space:
\beq
\delta(\k) \equiv {1\over{V}} \int d^3x \, \delta(\x) \, \exp{(i \k\cdot\x)} ,
\eeq
so that one can formally write, 
$\delta (\k) \equiv \sum_{n=1}^{\infty} \delta_n (\k)$, being 
$\delta_n (\k)$ the $n$-th order perturbative contribution. This
is expressed as an $n$-dimensional integral over the kernels, $F_n$, 
that encode the 
non-linear coupling of modes due to the gravitational evolution,
\bea
\delta_n(\k) &=& \int  d^3q_1 ... d^3q_1 \, \delta_D(\k-\q_1...-\q_n) 
\nn \\
&\times& F_n(\q_1...\q_n) 
\, \delta_1(\q_1)...\delta_1(\q_n)
\label{deltan}
\eea 
where $\delta_D$ is the
Dirac function and the kernels 
$F_n$ are given by symmetric homogeneous functions of $\q_n$
with degree zero, that is, some geometrical average (see Fry 1984, 
Goroff \etal 1986, Jain \& Bertschinger 1994,
SF96a). 
In particular, for the second order we have:
\beq
F_2(\q_1,\q_2) = {5\over{7}} + {1\over{2}} {\q_1\cdot\q_2\over{q_1q_2}}
\left({q_1\over q_2}+{q_2\over q_1}\right)
+{2\over{7}}  \left({\q_1\cdot\q_2\over{q_1q_2}}\right)^2 
\label{f2}
\eeq
which reproduces Eq.[\ref{secpert}] in Fourier space.

\subsection{The Monopole Approximation}
\label{sec:monopole}

If we are in a situation where there is spherical symmetry 
we can substitute the kernels $F_n$ in Eq.[\ref{deltan}] by
numbers $\nu_n$.
This numbers are given by the {\em monopole} term in an expansion of the 
kernels in spherical harmonics as they correspond to their angle averages
(the outcome of the angular integrals in Eq.[\ref{gaustree}]). In other words,
if we decompose the kernels, $F_n$ in multipoles,
\beq
F_n(\k_1,...,\k_n) = F_n^{l=0} + \sum_{l=1}^{\infty}\,F_n^l ,
\label{harmonics}
\eeq
only the first term $F_n^{l=0}$, the {\em monopole} (angle average of $F_n$) 
which is the {\em spherically 
symmetric} contribution to the kernels, contributes to the Gaussian tree-level.
We shall define, for convenience, the {\em monopole} in terms of 
some numbers $\nu_n$ in the way,  
\beq
F_n^{l=0} \equiv <F_n> = \nu_n/n! ,
\label{monop}
\eeq
where the factor $n!$ reflects the fact that we are looking for some symmetric
amplitudes. For instance, from Eq.[\ref{f2}] we have,
\beq
F_2^{l=0} \equiv <F_2> = \nu_2/2! = 34/21.
\label{nu2}
\eeq
On the other hand,  we can write the density fluctuation as formed of   
some generic {\em local}
and {\em non-local} components,
\beq
\delta(\x) \equiv \delta^{loc}(\x) + \delta^{nloc}(\x)
\eeq
where {\em local} means that the evolved  
(non-linear) density contrast at a given point
is only a transformation of the linear density contrast
at the same point, what we shall call a {\em local-density} transformation,
$\delta^{loc}(\x)=\La[\delta_1(\x)]$.
The {\em non-local} component
gives the contribution from density fluctuations at any other points.
This
is non-vanishing to all orders beyond the second in PT as commented above 
(see \S\ref{subsec:PT}) since gravity induces a non-local (\ie
non-isolated) evolution of the density fluctuations.  
The equivalent to this {\it local-density} 
transformation in Fourier space would then be,  
$\delta^{loc} (\k) = \La[\delta_1 (\k)]$. In particular, we can write the 
$n$-th perturbative order of $\delta^{loc} (\k)$ in the following way,
\beq
\delta_n^{loc} (\k) = {c_n \over n!}\,\delta_l (\k) * \dots * \delta_l (\k)
\eeq
which involves $n$ convolutions of $\delta_l (\k)$. By substituting
$F_n$ by its monopole contribution, $\nu_n/n!$, in Eq.[\ref{deltan}],
we immediately 
see that the above $c_n$ numbers are just those given by the {\em monopole} 
term, \ie $c_n = \nu_n$. Transforming back to real space, we find 
$\delta_n^{loc} (\x) = {c_n/n!}\,[\delta_l (\x)]^n$, and thus that 
{\em the monopole approximation
is given by the local contribution to the density
fluctuation} and has the form,
\beq
\delta^{loc} (\x) \equiv \sum_{n=1}^{\infty}\,\delta_n^{loc} (\x) = 
\La[\delta_1(\x)]=
\sum_{n=1}^{\infty}\,{c_n \over n!}\,[\delta_l(\x)]^n ,
\label{local} 
\eeq 
with $c_1 =1$, to reproduce the linear solution.
 and $c_n=\nu_n$ to recover the monopole.

\section{Cumulants in Perturbation Theory and Other Approximations}
\label{sec:PT}

We will now show the connection between the tree-graphs in the
diagrammatic approach to the one-point statistics and the {\em monopole}
contribution to the kernels. This will allow us to use a simple
local non-linear transformation (what we shall call the {\em local-density} 
transformation) to derive the predictions for the cumulants in the  
perturbative regime from the {\em monopole} contribution alone. 

\subsection{Statistical Properties}

The perturbative solutions above can be used to study the evolved statistical
properties of the density fluctuations, which is our final goal. 
Here we concentrate on the $J$th-order moments of the 
fluctuating field:  $m_J \equiv \lexp \delta^J \rexp$.
In this notation the variance is defined as:
\beq
Var(\delta) \equiv \sigma^2 \equiv m_2 - m_1^2
\eeq
In general, it is interesting to introduce the {\it connected 
moments} $\xi_J$, which carry statistical information independent of the
lower order moments, and are formally denoted by a bracket with
 subscript $c$:
\beq
\xi_J \equiv \lexp \delta^J \rexpc
\eeq
The connected moments are also called {\it cumulants},
{\it reduced} moments or {\it irreducible} moments. 
For a Gaussian distribution $\xi_J=0$ for $J>2$.
In many cases the higher order cumulants of a particular
distribution are given in terms of the second order one $\xi_2$.
It is therefore convenient to introduce some more definitions.
A {\it dimensional scaling} of the higher order moments in terms of the second
order one $\xi_2=\sigma^2$ is given by the following ratios:
\beq
B_J \equiv {\xi_J \over{\sigma^J}} = {\xi_J \over{\xi_2^{J/2}}}
\label{dimratios}
\eeq
We will see that for Gaussian initial conditions in perturbation theory, 
it is more useful to introduce the {\it hierarchical coefficients},
\beq
S_{J}\,=\,\frac{\xi_J}{{\xi_2^{J-1}}},
\label{sj}
\eeq
as PT predicts this quantities to be time-independent quantities on 
large scales.
These amplitudes are also called normalized one-point cumulants or 
reduced cumulants.
We shall also use
{\it skewness}, for $S_3$ and {\it kurtosis}, for $S_4$. 



The way to proceed is to substitute
the PT calculations Eq.[\ref{deltan}] and commute the spatial integrals 
in $\delta_n$ with the expectation values $\lexp ... \rexp$, by using
the {\it fair sample hypothesis} (\S 30 Peebles 1980). 
Thus the calculation will
reduce to averages over moments of $\delta_1$ with the kernels in 
Eq.[\ref{deltan}], which will, in turn, give us the statistical properties
of $\delta$ in term of the ones in the IC.

\subsection{Linear Theory}

If we only consider the first term in the PT series, Eq.[\ref{pt}],
and the growing mode, Eq.[\ref{Dt}], the cumulants of
the evolved field will just be,
$\lexp \delta^J \rexpc = D^J ~ \lexp \delta_0^J \rexpc$,
were $\lexp \delta_0^J \rexpc$ correspond to the cumulants in the IC.
Consistently, the hierarchical ratios (see Eq.[\ref{sj}]) will scale as,
$S_J = {S_J(0)/{D^{J-2}}}$,
were $S_J(0)$ are the initial ratios.
Note that this implies that the linear growth erases the initial
hierarchical ratios, so that $S_J \rightarrow 0$, as time evolves (and $D
\rightarrow \infty$). 

In terms of the dimensional scaling, see Eq.[\ref{dimratios}], we have,
$B_J = B_J(0)$, so that the linear growth preserves the initial values.
Note that if we want to do a meaningful calculation of these
ratios or the cumulants, in general, we might need to consider more
terms in the perturbation series, Eq.[\ref{pt}], depending on 
the statistical properties of the IC, \eg how they scale
with the initial variance, which is typically the smallness
parameter in the expansion of the cumulants.

For GIC both $B_J(0) = S_J(0) = 0$,
and we have to consider higher order terms in the perturbation
series to be able to make a non-vanishing prediction.

\subsection{The Tree-Level and Tree-Graphs}
\label{sec:gausstree}



The computation of the cumulants in PT dates back to Peebles (1980) work
where the leading order contribution to the skewness was obtained making
use of the second-order PT as given in Eq.[\ref{secpert}] to give,   
\beq
S_3 \equiv {\lexp \delta^3 \rexpc\over{\lexp \delta^2 \rexp^2}}
= {34\over{7}}  + \Or[\sigma_l^2],
\label{peebles}
\eeq
in agreement with the hierarchical scaling: $\xibar_J \simeq \xibar_2^{J-1}$.
Fry (1984) extended Peebles analysis by making the connection between tree
diagrams (or tree-graphs)
and the perturbative contributions to leading order in the 
Gaussian case. With the help of this formalism
he was able to obtain the leading order contributions for the 
three- and four-point 
functions making use of the 2nd and 3rd order in PT. Later on, 
Bernardeau (1992) found the generating function of the one-point cumulants 
to leading order for GIC. Furthermore, Fry (1984) found that, in general,
the lowest order (tree-level)
connected part that contributes to the $J$-order cumulant, $\xi_J$,
is of order $2(J-1)$ in $\delta_1$
from terms like:
\beq
\lexp \delta_1^2 \delta_2^{J-2} \rexp + ...+  \lexp \delta_1^{J-1} 
\delta_{J-1} \rexp.
\label{tree}
\eeq
Note that this involves the cancellation of $J-2$ contributions to the 
connected moment of order $J$.
This is a property of the Gaussian initial conditions
for which all $\lexp\delta_1^J \rexpc$ vanish 
for $J>2$. 
The above result Eq.[\ref{tree}] can be understood as follows.
We are only interested in estimating the leading-order contribution
(in terms of $\delta_1$) to the connected part of the
spatial average of the product $\delta(\k_1) \delta(\k_2)...\delta(\k_n)$,
\eg that contribution made with terms that can not be factorized 
as products containing
disjoint subsets of the labels. Each $\delta$ is to be replaced with
the perturbative expansion (see Eq.[\ref{pt}]), so that it may contain an
arbitrary power in $\delta_1$.  
In the Gaussian case, the moments
factorized only by pairs, so that the smaller contribution, 
with the smaller number of $\delta_1$, connecting $J$ points
has $J-1$ pairs. Each pair corresponds to two $\delta_1$ and
therefore the leading contribution has indeed $2(J-1)$ terms.
These are the tree diagrams, corresponding to graphs with no
loops (as any graph involving loops implies higher orders
in $\delta_1$). Therefore we see that for GIC the leading-order 
contribution to the cumulants 
(the so-called tree-level)
is only made of tree diagrams (or tree graphs). That is why it is called
the {\em tree-level}. We shall see in a companion paper (see Paper II) 
that the leading-order to the cumulants for non-Gaussian IC is not
solely made of tree diagrams as there are additional terms which depend
on the initial $n$-point correlation functions.

\subsection{The Tree-Level and
the Monopole Approximation}
\label{sec:local}

Consider a generic term of a tree-level contribution
such as $\lexp \delta_n(k_1) \delta_1(k_2)...
\delta_1(k_{n-1}) \rexp$. From Eq.[\ref{deltan}] we have:
\bea
&& \lexp \delta_n(\k_1) \delta_1(\k_2) \dots  \delta_1(\k_{n-1}) \rexp
= \int  d^3q_1 ... d^3q_n  \nn \\
&\times & \delta_D(\k_1-\q_1...-\q_n) \, F_n(\q_1...\q_n)  \nn \\
&\times &
\lexp \delta_1(\q_1)...\delta_1(\q_n) \delta_1(\k_2)...\delta_1(\k_{n-1}) \rexp
\label{deltani}
\eea 
Thus we only have to perform the spatial integral over the initial
moments  $\lexp \delta_1^{2(n-1)} \rexp$. In the case of 
GIC these moments are just products of two-point
functions. The only terms in the products of two-point functions
that produce {\it connected} graphs have pairs connected like:
\bea
&& \int  d^3q_1 ... d^3q_n \, \delta_D(\k_1-\q_1...-\q_n) \, F_n(\q_1...\q_n) 
 \\
&& \times \lexp  \delta_1(\q_1) \delta_1(\k_2)\rexp 
\lexp \delta_1(\q_2) \delta_1(\k_3)\rexp 
\dots \lexp \delta_1(\q_n) \delta_1(\k_{n-1})\rexp, \nn
\label{gaustree}
\eea
\eg those terms with pairs of $\q$ and $\k$, as otherwise there 
is a part of the 
integral that factorizes.
By isotropy, one sees that the only dependence in the 
angles between
 the $\q$'s comes through $F_n(\q_1...\q_n)$. Thus,
the geometrical dependence  in Eq.[\ref{gaustree}] can be
integrated out and the contribution of the kernels $F_n$ just becomes 
a constant.
Thus,  {\em the (Gaussian)
tree-level is given by the monopole and therefore by the local contribution 
to the density fluctuation} (see \S \ref{sec:monopole}).
 As a result, in this scheme,
the higher-order multipoles in the kernels 
are only expected to yield a non-vanishing
contribution in the next-to-leading orders (loops) in the cumulants, while
the monopole  contributes to
all perturbative orders. In what follows we shall
deal with the {\em local} contribution to the density fluctuation alone, 
$\delta_{loc}$ (monopole in the kernels) and 
work out its 1-point statistics. 

\subsection{Estimation of the Cumulants}
\label{sec:estimasj}

We can now easily estimate 
{\it all} the 1-point statistical properties in the {\em monopole} 
approximation to PT.
This can be done by using the
generating function method:
\beq
\xi_J = \lexp\delta^J \rexpc = \left. {d^J \ln[Z] \over{dt^J}}\right|_{t=0} , 
\label{generating}
\eeq
where $Z = \lexp e^{t \delta} \rexp$ is the {\em moment generating 
functional} of the one-point probability distribution functional,
and the field $\delta \equiv \delta_{loc}$ is given by Eq.[\ref{local}].
Furthermore, $\ln[Z]$ is the {\em cumulant generating functional} so that
Eq.[\ref{generating}] yields a simple derivation of the cumulants 
in terms of the moments.
A general discussion in terms of the
$p$-point correlation functions and its computation using
diagrammatic techniques (similar to those in Quantum Field Theory) is
given in detail in SF96a. 

The resulting expressions can be found
in Fry \& Gazta\~naga (1993, FG93 hereafter), 
who consider a generic local transformation
between density fields and find, to leading terms in $\sigma_l$:
\bea
S_{3} &=& 3 c_2 + \Or(\sigma_l^2) \nn \\
S_{4} &=& 4 c_3 + 12 c_2^2 + \Or(\sigma_l^2) \nn \\
S_{5} &=& 5 c_4 + 60 c_3 c_2+ 60 c_2^3 + \Or(\sigma_l^2)
\label{hierarg}
\eea
for Gaussian initial conditions.
Note that the leading contribution to $S_J$ is enough
to specify all the coefficients $c_n$ in the transformation, and therefore
all higher order corrections.
As expected, by substituting $c_2$ by the monopole contribution:
$c_2=\nu_2$ in 
Eq.[\ref{nu2}],
one recovers Peebles (1980) calculation, Eq.[\ref{peebles}].

\subsection[The Local-Density Approximation]
{The Local-Density Approximation and
the Equations of Motion.} 
\label{sec:eqofla}

We shall stress that the {\em local-density transformation} 
(see Eq.[\ref{local}])
is just a particular case of what
is usually understood as {\it local}. In general, it means that
the value of the evolved
field at a given point is a transformation of the 
IC at the same point, say, the initial value of the
field, its derivatives and any other fields (\eg the velocity field)
contributing at that very point.
Although gravity is non-local in this restricted sense,  
the above results show that
the statistical average of the non-local integrals 
involved in the reduced moments 
to leading order are exactly given by a {\em local-density} transformation.

However, to estimate the leading order (tree-level) it is  not
necessary to calculate the full kernels $F_n$, as we only need the 
$c_n$ numbers,
the {\it monopole} (spherically symmetric) contribution.
Therefore, {\em the values of $c_n$ (and thus the appropriate local-density 
transformation)
can be determined by finding
the spherically symmetric solution to the equations of motion}. 
We shall see below that for the gravitational evolution of the density field
in Lagrangian space,
this transformation is given by the spherical
collapse (SC) model, whereas for the ZA to the dynamics, 
it is given by what we shall call the Spherically
Symmetric ZA (SSZA).

\subsection{Other Approximations}
\label{sec:other}

 In one dimension, the probability distribution 
function of the matter field may be integrated in 
Lagrangian coordinates making use of the Zel'dovich Approximation (ZA) which 
recovers the exact dynamics. For the three-dimensional case however, the
ZA fails to describe the exact picture since the displacement field (that
only depends on the IC) no longer
factorizes in the actual solution for the growing mode as the ZA assumes 
(Bernardeau $\&$ Kofman 1995).

In the context of Lagrangian space and inspired by the
successful results of the Zel'dovich Approximation (hereafter ZA), 
there have appeared in the last years a number of papers
presenting some local approximations to the exact dynamics that
successfully describe the evolution of the density contrast in its first 
non-linear stages (see 
Bertschinger and Jain 1994, Hui
and Bertschinger 1996).
Most recently, Protogeros and Scherrer (1997, hereafter PS97), following the
formalism of the ZA, made use of a closed analytic expression,
$\delta = (1\,-\,\delta_l/\alpha)^{-\alpha}\,-\,1$,
to derive the cumulants following the formal analogy between
the density contrast and the vertex generating function at tree level
(see B92), 
for different values of $\alpha$, which
they take to be different approximations to the exact PT. For the 
particular case when $\alpha=3$ (what they call the ``spherical
approximation'') their approximation happens to be equal to the 
spherically symmetric solution to the
dynamics (in 3 dimensions) for the ZA (what
we call SSZA in Appendix \ref{app:za}). They also consider the case
$\alpha = 3/2$ as another approximation they call the ``exact
approximation'' (which, in fact, only reproduces the
exact tree-level  when $\Omega = 0$).
In general, none of these 
local approximations recovers the exact tree-level amplitudes as predicted
by PT (which we will show later that is given by the SC model).  

 On the other hand, a number of non-local approximations
to the non-linear dynamics of the cosmological fields have been designed
as to simplify the equations of motion by linearising any of the
fields involved.
The Linear (or Frozen) Potential Approximation (LPA) which assumes
that the gravitational potential generated by the density perturbations 
remains
linear throughout the evolution (see Brainerd \etal 1993, Bagla and
Padmanabhan 1994), 
the Frozen Flow Approximation (FFA) that takes the velocity field
to be frozen to its initial shape (see 
Matarrese \etal 1992), along with the ZA, which deals with
the approximation that particles move on straight lines in the
comoving picture (see Zel'dovich 1970), are the best known among them.
A general treatment in terms of the vertex generating function can be
found in Munshi \etal (1995). 
For a review of all these approximations see Bernardeau \etal (1994).
If we write the second perturbative order as given by
Eq.[\ref{secpert}], in the generic form:
\beq
\delta = 
\delta_l\,+A\,\delta_l^2\,+\,B\,\delta_{l,\alpha}
\,\Delta_{,\alpha}\,+\,C\,\Delta_{,\alpha \beta}\,
\Delta_{,\alpha \beta},
\eeq
one may derive the skewness at tree-level in the parametric fashion,
\beq
S_3 = 9\,A-3\,B\,+5\,C ,
\eeq
from which we can characterize the different above cited approximations 
depending on the 
weight each of them gives to these three different contributions:
while PT gives $A=5/7$, $B=1$, $C=2/7$, the ZA yields $1/2$, $1$, $1/2$,
whereas the FFA gives $1/2$, $1/2$, $0$, and the
LPA gets $1/2$, $7/10$, $1/5$, as the corresponding values for $A$, $B$ and
$C$ respectively, in each approximation. By just replacing the numbers in $S_3$
one sees that none of these non-linear approximations is able to reproduce 
the exact tree-level amplitude given by PT for the 
skewness. This illustrates that, despite successfully describing the 
first non-linear stages of the 
gravitational collapse of a density fluctuation,
none of the above given approximations 
yields to accurate enough predictions concerning the 
one-point cumulants
in the perturbative regime. 

\subsection{Higher Perturbative Orders}

Whenever $\delta \simlt 1$, corrective terms beyond  
leading order
(tree-level) become quantitatively important and must be 
taken into account. 
A systematic
approach for deriving the next-to-leading orders 
in PT was
introduced by  Goroff \etal (1986) by representing them in terms of  
the {\it loop} diagrams. Recently, the first results for the
{\it loop} corrections in PT have been calculated (see
SF96a, SF96b and S97).

The angle averaged (spherically symmetric) picture described above can 
be used to estimate the {\em monopole} contribution 
to an arbitrary higher-order in PT. This is easy to do as
one just has to take more terms in the series given by Eq.[\ref{local}] to
calculate the next orders in the moments.
As expected, even in the Gaussian case,
the higher-order terms coming from the monopole contribution differ 
from the exact
PT estimates. 
Bernardeau (1994c, B94c hereafter) already realised that fact 
when tracing the non-linear evolution of a 
volume occupied by particles that were in a given initial density perturbation.
If one assumes that the matter content of a fluctuation is conserved during
its collapse, then tracing the evolution of the volume is equivalent to
following that of the matter density. In his work, the
density fluctuation follows the SC model in what he called the 
{\em rare event limit}, \ie when $\delta_l/\sigma_l \rightarrow \infty$,
thus it is only exact as a mean picture of the actual collapse, 
with no scatter, i.e. $\sigma_l = 0$.

We have seen above that, in Fourier space, the
departure of the SC picture from the exact PT
is because the exact integrations
in this case, \eg  such as that in Eq.[\ref{deltani}] with
more $\delta_l$ factors, involve loops and the
geometrical dependence is not trivial anymore (the dipole, quadrupole...
contributions no longer cancel); The kernel $F_n$ integration will give
a different number depending on the loop configuration  (see SF96a),
contrary to the monopole 
approximation (were $F_n$ is always replaced by the same amplitude $c_n/n!$).
However, one can naively expect 
the monopole contribution to dominate by symmetry
because asymmetric contributions (arising from tidal terms) 
tend to cancel when averaged. 
In this paper
we shall compute the monopole contribution to PT and will compare it to the
exact analytic results, when available, or N-body simulations, to see how 
accurate this approximation is.

\section{The Spherical Collapse Model}
\label{sec:sc}

\subsection{The Shearless Approximation in Lagrangian Space}
\label{lagrsc}

We next want to derive the evolution of a density fluctuation
within the spherical model of collapse. In order to do that, we turn to
the Lagrangian space which is the natural framework for describing
the motion of a fluid element. The SC dynamics is
fully described in terms of one parameter, the initial size of the
spherical fluctuation. In this sense, the SC model is just a particular case
within the family of {\it local-density} approximations discussed above. 
When we write the equations in Lagrangian space,
the natural variable is the density contrast and the only parameter
in terms of which the solution is given is the linear density
contrast. 

 We first must define the
conformal time $\tau$ which is the comoving time parameter to the 
motion of the mass  and we shall denote with a dot
the associated time derivatives (conformal time derivatives), defined as,
\beq
{d \over d\tau}\,=\,{\partial
\over \partial\tau}\,+
\,\vv\cdot\nabla_q \equiv a(t){d \over dt}, 
\eeq
$a(t)$ being the scale factor.
The generic equation describing the 
evolution of a spherically symmetric perturbation in an expanding 
universe is given by,
\beq
{d^2 R \over dt^2}\,=\,-{G\,M(R) \over R^2} = 
-4\pi G \rho R
\label{scradial}
\eeq
since the matter contained in a spherical perturbation of radius R is, 
$M(R) = 4\pi \rho R^3/3$. 
Therefore, the spherical density perturbation 
$\delta = (R/a(t)R_0)^{-3}-1$, is described by the following equation of 
motion
\beq
{\ddot{\delta}}\,+\,{\cal H} (\tau)\,{\dot{\delta}}\,
-\,{4\over 3}\,{{\dot{\delta}}^2 \over (1+\delta)}\,=\,
4\pi G (\rho-\rhobar) a(\tau)^2 (1+\delta).
\label{scdyn}
\eeq    
where the dot denotes a conformal time derivative, and $a(\tau)$ is the
scale factor in terms of the conformal time from which we define the 
conformal Hubble parameter ${\cal H}(\tau) = d \log a(\tau)/d\tau$.
Introducing the comoving time derivative, $d/dt \equiv \cdot$, the last 
equation translates into,
\beq
{\ddot{\delta}}\,+2\,H (t)\,{\dot{\delta}}\,-\,{4\over 3}\,{{\dot{\delta}}^2 
\over (1+\delta)}\,=\,
4\pi G \rhobar\, \delta \,(1+\delta) ,
\label{comsc}
\eeq     
where $H (t) = d \log a(t)/dt$.

On the other hand, the Newtonian Fluid equations with zero pressure, 
which are the appropriate 
description for the sub-horizon modes in a perturbed FRW universe, 
in its matter 
dominated regime (the relevant for describing formation of structures),
are usually given 
in Euler space (see \eg \S 9 in Peebles 1980). If we now turn 
to {\em Lagrangian} coordinates 
and  we re-express the equations of motion in terms of the 
derivatives of the conformal time $\tau$,
the continuity equation reads,
\beq
{\dot{\delta}}\,+\,(1+\delta)\,\theta\,= 0, 
\quad \theta\equiv\,\nabla\cdot\vv .
\label{conteq}
\eeq
We then  combine this last expression with the Raychaudhuri equation,
\beq
{\dot{\theta}}\,+\,{\cal H}(\tau)\,\theta\,+
\,{1\over 3}\,\theta^2\,+\,\sigma^{ij}
\sigma_{ij}\,-\,2\,\omega^2 = -4\pi G \rhobar \delta a^2
\eeq
where
$\omega^2\equiv {1\over 2}\,\omega^{ij}\omega_{ij}$, and
the expansion $\theta$, vorticity $\omega_{ij}$, and shear $\sigma_{ij}$,
are given  by the trace, traceless
antisymmetric and symmetric parts respectively, of the velocity divergence,
\bea
\nabla_i v_j &=& {1\over 3}\theta\delta_{ij}\,+\sigma_{ij}\,+\,
\omega_{ij}, \nn \\
&& \sigma_{ij} = \sigma_{ji}, \quad \omega_{ij} = -\omega_{ji},
\eea
and get a second-order differential equation for the density contrast,
\bea
&& {\ddot{\delta}}\,+\,{\cal H}(\tau)\,{\dot{\delta}}\,-\,{4\over 3}\,
{{\dot{\delta}}^2 
\over (1+\delta)} \nn \\ 
&& \,=\, (1+\delta)\left(\sigma^{ij}
\sigma_{ij}\,-\,2\,\omega^2\,+\, 4\pi G \rhobar\delta a^2  \right).
\label{gravinst}
\eea  
For an initially irrotational fluid (the 
expansion preserving its
irrotational character in the linear regime): $\omega =0$. 
Making the further assumption that 
there is {\em no shear}, Eq.[\ref{gravinst}] leads to the equation we obtained 
for the SC model (see Eq.[\ref{scdyn}] above). In other words, {\em the SC 
approximation is the actual dynamics when tidal
effects are neglected}. As one would expect, this yields a
{\em local} evolution, in the restricted sense that the evolved field 
at a point is just given by a local (non-linear)
transformation of the initial field at the same point.
Throughout this paper we shall drop the shear term and work out 
the solution to this dynamics in the perturbative regime.

 The exact (non-perturbative) solution for the SC of the density contrast 
in an Einstein-deSitter universe admits a well-known parametric representation,
\bea
\delta(\phi) = {9 \over 2}{(\phi-\sin\phi)^2 \over
(1-\cos\phi)^3} - 1, \quad \delta_l (\phi) = {3\over 5} 
\left[{3\over 4}(\phi-\sin\phi)\right]^{2/3}  \nn 
\eea
for $\delta_l > 0$, linear overdensity, and
\bea
\delta(\phi) = {9 \over 2}{(\sinh\phi-\phi)^2 \over
(\cosh\phi-1)^3} - 1, \quad
\delta_l (\phi) = - {3\over 5} \left[{3\over 4}
(\sinh\phi-\phi)\right]^{2/3}, \nn
\eea
for $\delta_l < 0$, linear underdensity (see Peebles 1980), where
the parameter $\phi$ is just a parameterization of the time coordinate.  

If we are only interested in the perturbative regime 
($\delta_l \rightarrow 0$), which is the relevant one for the description of
structure formation on large scales,
the above solution can be also expressed directly
in terms of the initial density contrast, which plays the role
of the initial size of the spherical fluctuation in 
Eq.[\ref{comsc}]. This way,
 the evolved density contrast in the perturbative regime is given 
by a {\em local-density}  transformation of the 
linear density fluctuation,
\beq
\delta  =  f(\delta_l) =  
\,\sum_{n=1}^{\infty} {\nu_n \over n!}\, {[\delta_l]^n}
\label{loclag}
\eeq
Notice that all the dynamical information in the SC model is encoded 
in the $\nu_n$ coefficients of this {\em local-density 
transformation}, Eq.[\ref{loclag}].
In an Einstein-de Sitter universe ($\Omega =1, \Lambda=0$ FRW universe), 
we can introduce the above power series expansion in Eq.[\ref{comsc}] 
and determine the 
$\nu_n$ coefficients one by one. The first ones turn out to be,
\bea
&&\nu_2 = {34\over 21} \sim 1.62; \quad
\nu_3 = {682\over 189} \sim 3.61 \nn \\ 
&&\nu_4 = {446440\over 43659} \sim 10.22; \quad
\nu_5 = {8546480\over 243243} \sim 35.13 
\label{nusc}
\eea
and so on. 

Remember that this evolution is in {\em Lagrangian} space,
$\delta=\delta(\q)$.
We would like to relate the above results, obtained in 
Lagrangian coordinates  to the corresponding fluctuation in 
Eulerian coordinates. In Lagrangian space a fluid element of
a given mass 
is labeled by its initial position $\q$ (or Lagrangian coordinate), 
whereas Eulerian space uses the density $\rho(x)$ at the
final coordinate $\x=\x(\q)$.
Notice that mass conservation requires that the volume elements be related 
like: $d^3\q=(1+\delta) d^3\x$. Thus density probabilities in
Lagrangian ($_L$) and Eulerian ($_E$)
space  should
also be related by the same factor, 
$\Delta_L \delta = (1+\delta) \Delta_E \delta$ 
 (see Kofman \etal 1994).
This provides a simple way 
to translate density moments in Eulerian space, $\lexp (1+\delta)^J \rexp_E$
with the ones in Lagrangian space $\lexp (1+\delta)^J \rexp_L$,
\bea
&& \langle (1+\delta)^J \rangle_L = \int (1+\delta)^J \,
P[\delta] \, \Delta_L \delta  = \nn \\
&& \int  (1+\delta)^{J+1} 
P[\delta] \, \Delta_E \delta 
= \langle (1+\delta)^{J+1} 
\rangle_E.
\label{lag2eul+}
\eea
In particular, as pointed out by Bernardeau (B94b)
(and also PS97), the conservation of Eulerian volume yields,
$\langle 1 \rangle_E = 1 = \langle (1+\delta)^{-1} \rangle_L$,
which requires a normalization for $\delta$ in Eq.[\ref{loclag}]:
\beq
1+\delta = (1 + f) \, \langle (1+f)^{-1} \rangle_L.
\label{normalag}
\eeq
It is interesting to note 
from Eq.[\ref{lag2eul+}] that:
\beq
\langle \delta^J \rangle_L = \langle \delta^J \rangle_E 
+ \langle \delta^{J+1} \rangle_E .
\label{lag2eul}
\eeq
Thus, to leading order, there is no difference between
Eulerian and Lagrangian moments. In general, the leading 
contribution to the cumulants is enough
to specify all the coefficients $c_n$ (see \eg Eq.[\ref{hierarg}]), of 
the local transformation provided by Eq.[\ref{local}]. This means that the local-density
transformations that produce the tree-level in PT are identical in Euler
and Lagrangian space. A similar result was noted by
Protogeros \& Scherrer (PS97), in the context of
hierarchical distributions, but this  argument is more general 
as applies to any non-Gaussian distributions where:
$ \langle \delta^{J+1} \rangle_E < \langle \delta^J \rangle_E$,
which holds even for the strongly non-Gaussian dimensional models,
Eq.[\ref{dimratios}], that we shall investigate later 
(see Paper II).
These models for the IC
naturally appear as solutions for the topological defects models
for structure formation. 

\subsection{The SC as a Solution to the Cumulants}
\label{sec:sccum}

Our goal here is to relate the cumulants of the initial distribution
with the ones resulting from the full  non-linear evolution of the field.
To estimate the evolved cumulants under the SC model, 
we perform the following prescription:

\begin{itemize}

\item (a) we start with the initial cumulants, in the limit
of small fluctuations, $\delta\rightarrow 0$ .
In this limit, the cumulants are equal 
in Lagrangian and Eulerian space, as $d^3\q \rightarrow 
d^3\x$.

\item (b) we then use the local transformation Eq.[\ref{loclag}], 
with its proper
normalization, Eq.[\ref{normalag}],
to relate the initial and final cumulants in
Lagrangian space. As argued above, this is equivalent to include the
contribution from the monopole alone.
This can be done as described in section \S\ref{sec:estimasj}
using Eq.[\ref{generating}] to obtain expressions such as Eq.[\ref{hierarg}]
 by keeping the relevant terms. 

\item (c) finally we use Eq.[\ref{lag2eul+}] to relate the Lagrangian 
and Eulerian moments, which can be rewritten in a more compact way:
\beq
\langle \delta^J \rangle_E =\langle \delta^{J-1} \rangle_L -\langle \delta^{J-2} \rangle_L+ \langle \delta^{J-3} \rangle_L - \dots   + \, \langle \delta \rangle_L,
\label{eul2lag}
\eeq
from which we get the cumulants in Euler (real) space.
\end{itemize}
Note that the cumulants obtained from (b)+(c), which
are Appendix \ref{app:locumul} (see Eq.[\ref{scgic}]), 
are not the same as the
ones obtained using the local-density relation in 
{\em Euler} space directly (see Appendix \ref{app:euler}).
As explicitly seen from Eq.[\ref{lag2eul}], they are 
only the same to leading order. We choose this approach 
because the Lagrangian space, comoving with the fluid element,
is the natural coordinate system for a local description of its evolution.
Note that a local-density description in Lagrangian space will in general 
mean a non-local one in Euler space.

\subsection{The SC Model and PT}
\label{sec:tree}

We have pointed out in the previous sections that
the monopole contribution (which is the exact
result for tree-graphs)
to the cumulants in PT is given by a local-density
transformation Eq.[\ref{local}],
whose coefficients, $c_n$, 
are to be determined by the kernels, $F_n$ (see Eq.[\ref{deltan}]), under the 
relevant dynamics.  These arguments are valid for any dynamics and
apply either to Euler or Lagrangian space. They are true for any 
leading order calculation. 
As argued in section \S\ref{sec:eqofla},
to estimate this contribution to the cumulants it is  not
necessary to calculate the full kernels $F_n$, as we only need
the numbers $c_n$.
Given the equations for the evolution of a field,
one can determine Eq.[\ref{local}] and therefore $c_n$
by just requiring the solutions to be {\em spherically symmetric}.

We have shown above that for gravity, the spherically symmetric
solution to the evolution of density perturbations  
is given by Eq.[\ref{scdyn}], \eg the SC model, whose
solution is well-known (see Eq.[\ref{loclag}]). 
Thus, $\La = f$ which yields the monopole
contribution, $c_n=\nu_n$, without need of estimating the kernels
$F_n$ or any integral. Note, in particular, that $\nu_2=34/42$
from both approaches (see Eqs.[\ref{nu2}],[\ref{nusc}]).

This provides with a
simpler derivation and interpretation of the results presented
by B92, who found  the values $\nu_n$
that give the leading-order contribution to PT for
Gaussian initial conditions. In the language of  B92,
the vertex generating function, $\calG(-\tau) = f(\delta_l)$
(see also PS97). Our derivation explains therefore
why the vertex generating function $\calG(-\tau)$
follows the SC model, which lacked a satisfactory explanation in the
context of B92.
Besides its simplicity, 
in our framework one has the added advantage of being able to
use the local-density relation to estimate the higher-order 
corrections for both Gaussian and  non-Gaussian initial conditions
(see Paper II).
Note nevertheless that there is an important difference
in practice with the work of B92. His 
vertex generating function $\calG(-\tau)$, corresponds to 
cumulants in Euler space, while our local-density relation
$f(\delta_l)$, applies to Lagrangian space. To leading order,
both give identical results for Gaussian IC, but 
they yield different results in general for Non-Gaussian IC or
for next-to-leading terms for Gaussian IC.


 We  want to stress that the fact the SC model determines the
tree-level amplitudes
is neither a singular property of the Gaussian nature of the
IC nor an argument that has to do with the flat geometry of 
space. It is a general property that follows straightforwardly from the
{\em local} nature of the monopole contribution to PT (see section 
\S\ref{sec:eqofla}):
{\em
the tree-graphs (and the monopole contribution)
in PT are exactly given by the SC model regardless
of either the statistical nature of the initial conditions or the geometry
of the universe}.  
 We shall illustrate the above statement in the framework of B92
calculations by showing, in an accompanying paper (Fosalba \& Gazta\~{n}aga
1998, Paper III in this issue),
that the equations of motion
that govern the leading order amplitudes (for Gaussian initial conditions)
in a universe with arbitrary density parameter $\Omega$, 
are those of the SC dynamics.

\subsection{Smoothing Effects}
\label{sec:smooth}

So far we have worked out the statistics of the unsmoothed 
fields. There remains to be seen whether the local-density approximation
may be extended to the statistics of the spatially smoothed fields, 
which is essential if we are
to compare with N-body simulations and observations.
For that purpose,
we have to introduce the fluctuating field integrated over 
a {\em finite} volume. This volume is fixed by the size of 
the window function that acts as a 
filter on the unsmoothed field. 

Our goal is therefore to relate the {\it smoothed}
cumulants in the evolved distribution with the {\it smoothed}
cumulants in the initial one. The latter are {\it inputs} to our 
calculations and should tell us, 
\eg how the smoothed {\em rms} fluctuation
changes as a function of the smoothing radius: $\sigmahat=\sigmahat(R)$.

We will focus here in
the top-hat filter, defined as, 
\bea
W_{TH} (\x,R) &=& 1 \quad if \quad \x \leq R , \nn  
\eea
being zero otherwise, 
where $R$ is the smoothing radius.
Note that this filter is spherically symmetric so that the transformation
that gets from the unsmoothed to the smoothed evolved fields preserves
the spherical symmetry.

In the local-density picture of the spherical collapse, each fluctuation
is isolated from the others and evolves according to the value
of the initial amplitude alone. The statistics of the evolved field 
is just induced by the statistics of the initial one, which could
in general be non-Gaussian. In order to estimate the statistical properties
(one-point cumulants) we can therefore picture the initial spatial distribution 
as just made of a single spherical fluctuation. 
 Different realizations of this fluctuation could have different 
amplitudes or shapes in a proportion
given so as to match the statistics of the initial conditions
(see \S61 Peebles 1980 for examples). In our case this
is given in terms of the {\it unsmoothed} distribution, which corresponds
to a top-hat smoothing of radius $R_0 \rightarrow 0$.

The spatial smoothing only acts on the fluctuation by
changing its amplitude. We want, by construction, 
the new amplitude to be a function of
the smoothing radius, $R$. 
Thus, for the initial conditions $\delta_l$, the smoothed field is:
\beq 
\deltahat_l(R) = {\sigmahat_l(R)\over{\sigma_l(R_0)}} \delta_l
\label{sigmahatl}
\eeq
as smoothing does not change the statistics directly
(at least for a top-hat). 
A similar argument applies to the evolved field, which under the SC
model is just a local transformation of the initial fluctuations,
which only changes the local amplitude:
\beq
\deltahat = {\sigmahat\over{\sigma}} \, \delta
\label{sigmahat}
\eeq
where $\deltahat$ is the smoothed fluctuation and $\delta$ the
unsmoothed one. 
The statistics of $\deltahat$ will therefore be given by 
the unsmoothed statistics, $\delta$, which
are in turn given by the local-density relation Eq.[\ref{loclag}]:
$\deltahat \sim \delta = f[\delta_l]$.
Using Eq.[\ref{sigmahatl}]
we express
this as a function of the linear smoothed fluctuation $\deltahat_l$:
\beq
\deltahat[\deltahat_l] \sim f[\deltahat_l {\sigma_l\over{\sigmahat_l}}]
\label{deltahat}
\eeq
where $\sim$ just means that $\deltahat$ needs to be normalized,
as explained before (see Eq.[\ref{normalag}]).
Note that the input {\em rms} fluctuation is a function of the 
smoothing radius,
$R$, which also fixes the
amplitude of the final smoothed fluctuation. 
As the smoothing radius $R$ is, by construction, larger than the
unsmoothed one $R >> R_0 \rightarrow 0$, we have for the smoothed
fluctuation:
$\rhohat = {m_0\over{V}} \sim  m_0 \, R^{-3}$,
as there is no other smoothing than the sharp cut-off,
(note that this may not be the case for other filters, 
such as the Gaussian smoothing,
which changes the initial shape of the fluctuation and therefore
the final mass).
Thus the smoothed amplitude, $\sigmahat=\sigmahat[R]$, will be a function 
of $\deltahat$ through $R$. 
This together with Eq.[\ref{deltahat}] can be used to obtained
$\deltahat$ in a recursive way.

In the case of a 
power-law power spectrum $P(k) \sim k^{n}$, the smoothed variance
is also a power-law $\sigmahat_l \sim R^{\gamma/2}$, where
$\gamma= -(n+3)$. We then have,
$\sigmahat_l=\sigma_l \, 
(1+\deltahat)^{-\gamma/6}$.
Note that $\gamma=0$ reproduces the unsmoothed result.
Moreover, using Eq.[\ref{deltahat}], we find:
\beq
\deltahat[\deltahat_l] \sim f[\deltahat_l  (1+\deltahat)^{\gamma/6}]
\label{deltahat2}
\eeq
up to a normalization factor given by Eq.[\ref{normalag}].
Note that this final result as well as the general expression
Eq.[\ref{deltahat}], agrees with B94a arguments,
 based on the vertex generating function. However,
Eqs.[\ref{deltahat}],[\ref{deltahat2}] 
do not limit themselves to Gaussian 
initial conditions or the leading-order term. 
Here again, the
vertex generating function $\calG (-\tau)$, corresponds to 
cumulants in Euler space, while our local-density relation,
$f(\delta_l)$, applies to {\em Lagrangian} space. To leading order,
they both  give identical results with Gaussian IC, but 
they yield different results, in general, for Non-Gaussian IC or
for higher-order terms with Gaussian IC.

We will further write the 
smoothed density contrast (over a
smoothing scale $R$) in the following way:
\beq
\deltahat= \hat{f}(\deltahat_l) \equiv 
\sum_{k=1}^{\infty} {{\overline{\nu_k}} \over k!}\, 
\left[\deltahat_l \right]^k,
\label{deltahat3}
\eeq
where the $\overline{\nu_k}$ are a generalization of the unsmoothed 
coefficients. 
By Taylor-expanding Eq.[\ref{deltahat2}] one obtains,
\bea
\overline{\nu_2} &=& \nu_2 + {\gamma \over 3} \nn \\
\overline{\nu_3} &=&{1 \over 4}(-2 \,\gamma + \gamma^2 + 6 \,\gamma\, \nu_2 + 4 \,\nu_3) \nn \\
\overline{\nu_4} &=& {1 \over 27} (36 \,\gamma - 36 \,\gamma^2 
+ 8 \,\gamma^3 - 
108 \,\gamma \nu_2 + 72 \,\gamma^2 \,\nu_2 + \nn \\
&+&        54 \,\gamma \,\nu_2^2 + 72 \,\gamma \,\nu_3 + 27 \nu_4) ,
\label{nusm}
\eea  
and so on.

For an arbitrary power spectrum $P(k)$, the above results can be trivially
generalized using Eq.[\ref{deltahat}]. The results are given in 
Eq.[\ref{nlscgp}] of the Appendix \ref{app:locumul}.


In Appendix \ref{app:nu2} we present an alternative derivation of the
(top-hat) smoothing effects within the SC approximation, following the
kernels in Fourier space. There, one can explicitly see
how in the SC model, smoothing is a trivial operation
which only changes the $\nu_n$ coefficients but not the local
nature of the transformation from the linear to the
non-linear density field.

\section{Comparison of the SC model with exact PT and N-body Simulations}
\label{sec:comparison} 

In this section we start from the exact solution to the 
SC dynamics in an Einstein-de Sitter universe assuming the {\em rms} 
fluctuation $\sigma_l \simlt 1$. In that perturbative limit, we
carry out the calculation of the connected moments for the density. 
Results for the velocity field
are provided in Paper III.

We compare the SC predictions with 
those derived from the exact PT in the 
context of the diagrammatic approach. Although we know that the SC dynamics 
exactly reproduces the leading-order contribution to the cumulants,
higher order effects will in general be different. We have seen that the
difference is due to neglecting the shear and can therefore 
be attributed to non-local effects, which we will also call
 {\it tidal} effects.
Note that in our context {\it local} refers only to the density. 

As commented above, the 
results are derived in general for the smoothed fields for a top-hat
window function, the unsmoothed fields
being recovered just as the particular case $\gamma = 0$.
To simplify the expressions, we will assume that higher order
derivatives of the variance (see Appendix \ref{app:locumul}, 
Eq.[\ref{eq:gamma}] with $\gamma_p \simeq 0$ for $p>2$) can be neglected. 
This is a good approximation
for slowly varying $P(k)$, like CDM (see Scoccimarro 1998)
or the APM model, but
it is straightforward to take those higher order derivatives 
into account anyway.
We do take into account $\gamma_2$, which for loop corrections
could contribute up to $10\%$ on small scales.

We will also compare the SC results with different N-body simulations,
with parameters given in Table \ref{Nbody}.
Also given is the reference where
the details for a particular run can be found.
For each simulation, 
the cumulants $\xibar_J$ are estimated from counts in spherical
cells, as described in
BGE95, where more details about the estimation are given. 
For the later outputs the correlations are corrected from
the Poisson shot-noise (\eg BGE95). We
also apply the corrections due to the ZA transient (see Baugh \etal 1995)
as analitycally derived by Scoccimarro (1998).

\begin{table*}
\begin{center}
\caption[dummy]{Simulation parameters}
\label{tab:param}
\begin{tabular}{lllccccc}
\hline
\hline  
run names& $P(k)$ & $\Omega$-$\Lambda$ & number    & mesh    & $L_{box} $  & Reference  \\ 
            & & & of particles         &         & $(\mpc) $  \\ \hline
\\
SCDM (a)-(j) & $\Gamma=0.5$ CDM  & $\Omega=1$ \ $\Lambda=0$ 
 & $126^{3}$ & $128^{3}$ & 378 &  Gazta\~{n}aga \& Baugh 1995\\
LCDM (a)-(j) & $\Gamma=0.2$ CDM  & $\Omega=0.2$ \ $\Lambda=0.8$ 
 & $126^{3}$ & $128^{3}$ & 378 &  Gazta\~{n}aga \& Baugh 1995\\
APMPK1 (a)-(e)  &APM &   $\Omega=1$ \ $\Lambda=0$  & $126^{3}$ & $128^{3}$ & 400& Gazta\~{n}aga \& Baugh 1997 \\
APMPK2 (a)-(b)  &APM &   $\Omega=1$ \ $\Lambda=0$  & $200^{3}$ & $128^{3}$ & 600&  Baugh \& Gazta\~{n}aga 1996\\ 
\hline \\ 
\label{Nbody}
\end{tabular}
\end{center}
\end{table*}


Following the steps described in section \S\ref{sec:sccum}, 
we can derive now the smoothed one-point cumulants to
an arbitrary perturbative order for the SC dynamics. 
In this case we start from Gaussian initial conditions and
use the local transformation for the
local  smoothed density (see Eq.[\ref{deltahat3}])
with $c_k=\overline{\nu_k}$, as given in Eq.[\ref{nusm}],
to find the leading-order and  higher-order corrections for
the variance and the hierarchical amplitudes. 
In order to handle the perturbative expansions in the cumulants for both
GIC or NGIC in a general framework, we shall introduce the following notation,
\beq
\sigma^2 = \lexp \deltahat^2 \rexp = \sum_{i} {s_{2,i}\,\sigma_l^i}
\eeq
where $s_{2,1} \equiv 1$ and the subscript $i$ in the coefficients 
label the perturbative order. For GIC the odd terms in the perturbative
expansion vanish and the get for the variance,
\beq
\sigma^2 \equiv \sigma^2_G = \sigma_l^2 \,+\,s_{2,4}\,\sigma_l^4\,+
\,s_{2,6}\,\sigma_l^6\,+ \cdots 
\eeq
Note that our notation for GIC is equivalent to that of SF96a provided one
identifies $s_{2,2 i+2} \equiv s^{(i)}$ with $i=1,2, \cdots$.

On the other hand, for the hierarchical amplitudes we keep the above 
introduced notation with the added labeling of the order of the moment
$J$, that defines the $S_J$ coefficients, 
\beq
S_J \equiv
{\xibar_J\over \xibar_2^{J-1}}\, \equiv
{\lexp \deltahat^J \rexp_c \over \lexp \deltahat^2 \rexp^{J-1}}\,= 
\sum_{i} {S_{J,i}\,\sigma_l^i}
\eeq
which for GIC has non-vanishing contributions from the even terms alone,
\beq
S_J^G = S_{J,0} \,+\,S_{J,2}\,
\sigma_l^2\,+\,S_{J,4}\,\sigma_l^4\,+
\cdots
\eeq
with $S_{J,2 i} \equiv S_J^{(i)}$ (with $i=0,1,2, \cdots$) in SF96a notation. 

As mentioned before, it is usual to denote 
{\it tree-level} to the leading order contributions, $S_J^{(0)}$. Although
this is true for GIC, in general the leading order contributions do not
correspond to the tree-level in the diagrammatic approach 
(see Paper II).
However, for the sake of clarity, we shall maintain the notation $S_J^{(0)}$
(or $T_J^{(0)}$ for the velocity field, see Paper III) to denote the leading-order 
contribution.

With this notation the results for GIC
(in terms of the SC coefficients
$\nu_k$) are given in Appendix \ref{app:locumul}, Eq.[\ref{scgic}], 
with  $c_k=\nu_k$. In terms
of the smoothing index $\gamma = -(n+3)$, we have,
for a power-law power spectrum
$P(k) \sim k^n$, and a top-hat window:
\bea 
s_{2,4} &=& {1909\over 1323}\,+\,
{143\over 126}\,\gamma + 
{11\over 36}\,\gamma^2 \nn \\
s_{2,6} &=& {344439415\over 107270163}\,+\,{21651395\over 3667356}\,\gamma\,+\,
{408721\over 95256}\,\gamma^2 \,+ \nn \\
&+& \,{1651\over 1134}\,\gamma^3\,+\, 
     {127\over 648}\,\gamma^4
\nn \\
S_{3,0} &\equiv& S_3^{(0)} =  {34\over 7} + \gamma \nn \\
S_{3,2} &=&
{1026488\over 101871} + {12862\over 
1323}\,\gamma\, + \,{407\over 126}\,\gamma^2 \,+ \,{10\over 27}\,
\gamma^3 
\nn \\
S_{3,4} &=& 
{251978977148\over 5256237987} + 
{71492200235\over 750891141}\,\gamma 
      +  {138567091\over 1833678}\,\gamma^2  \nn \\
&+& {79295\over 2646}\,\gamma^3 
+ {2891\over 486}\,\gamma^4 + 
      {1841\over 3888}\,\gamma^5 
\nn \\
S_{4,0} &\equiv& S_4^{(0)} = {60712\over 1323} + {62\over 3}\,\gamma + 
{7\over 3}\,\gamma^2 \nn \\ 
S_{4,2} &=&  {22336534498\over 83432349} 
+ {42649448 \over 130977}\,\gamma
+\,{3571621 \over 23814}\,\gamma^2\, \nn \\
&+& {35047 \over 1134}\,\gamma^3 
+\, {1549 \over 648}\,\gamma^4 \nn \\ 
S_{4,4} &=& {126152927186426522\over 61923739724847} 
+ {69638296109567 \over 15768713961}\,\gamma   \nn \\
&+&  {8977285860007 \over 2252673423}\,\gamma^2 
+ {7018518515 \over 3667356} \,\gamma^3  \nn \\ 
&+&  {24548155 \over 47628}\,\gamma^4  
+ {668971 \over 9072}\,\gamma^5 + 
{102005 \over 23328}\,\gamma^6 
\nn \\
S_{5,0} &\equiv& S_5^{(0)} =  {200575880\over 305613} + 
{1847200\over 3969}\,\gamma + 
{6940 \over 63}\,\gamma^2 + {235 \over 27}\,\gamma^3 \nn \\
S_{5,2} &=& {38066642685488 \over 5256237987} + 
{8041429493780\over 750891141}\,\gamma+ 
{5828197535 \over 916839}\,\gamma^2 \nn \\ 
   &+& {45012655 \over 23814}\,\gamma^3 + {955895 \over 3402}\,\gamma^4 
 + {4052\over 243}\,\gamma^5 
\nn \\
S_{6,0} &\equiv& S_6^{(0)} = {351903409720 \over 27810783} + 
{3769596070 \over 305613}\,
\gamma + {17907475\over 3969}\,\gamma^2  \nn \\
&+& {138730 \over 189}\,\gamma^3
+ {1210\over 27}\,\gamma^4 \nn \\
S_{6,2} 
&=& {93347762463213320\over 421249930101} 
+ {2034255356621746 \over 5256237987}\,\gamma  \nn \\
       &+&  {211757079765188 \over 750891141}\,\gamma^2 +
 {301575001360 \over 2750517}\,\gamma^3  \nn \\ 
&+& {69888305 \over 2916}\,\gamma^4 + {1582214 \over 567}\,\gamma^5 + 
{14591 \over 108}\,\gamma^6 
\eea

\begin{table}

\begin{center}

\begin{tabular}{|c||c|c|c|c|}
\hline \hline
SC & Unsmoothed & \multicolumn{3}{c|}{Smoothed} \\ 
\hline \hline
& $\gamma=0$ & $\gamma=-1$ & $\gamma=-2$ & $\gamma=-3$ \\ \hline
& $n=-3$ & $n=-2$ & $n=-1$ & $n=0$  \\ 
\hline \hline
$s_{2,4}$   & 1.44 & 0.61 & 0.40 &  0.79 \\ 
\hline
$s_{2,6}$   & 3.21 & 0.34 & 0.05 &  0.68 \\ 
\hline \hline
$S_{3,0}$ & 4.86 & 3.86 & 2.86 & 1.86 \\ 
\hline
$S_{3,2}$ & 10.08 & 3.21 & 0.59 & -0.02 \\ 
\hline
$S_{3,4}$ & 47.94 & 3.80 & 0.07 & 0.06  \\ 
\hline \hline
$S_{4,0}$ & 45.89 & 27.56 & 13.89 & 4.89  \\ 
\hline
$S_{4,2}$ & 267.72 & 63.56 & 7.39 & -0.16 \\ 
\hline
$S_{4,4}$ & 2037.2 & 138.43 & 1.99 & 0.31 \\ 
\hline \hline
$S_{5,0}$ & 656.31 & 292.35 & 96.50 & 16.52 \\ 
\hline
$S_{5,2}$ & 7242.2 & 1263.97 & 91.85 & -1.16 \\ 
\hline
$S_{5,4}$ & 80903.0 &  4363.92 &  42.89 &  1.53 \\
\hline \hline
$S_{6,0}$ & 12653.49 & 4141.58 & 876.62 & 67.81 \\ 
\hline
$S_{6,2}$ & 221597.1 & 28256.19 & 1274.38 & -8.04 \\ 
\hline
$S_{6,4}$ & 3405857.8 &  140641.3 &  906.74 &  7.88 \\
\hline \hline 

\end{tabular}

\caption[junk]{Values for the higher-order perturbative contributions in the
SC model for the unsmoothed ($n=-3$) and 
smoothed ($n=-2,-1,0$) density fields, for a top-hat filter an a power-law
power spectrum.}
\label{nlsc}
\end{center}

\end{table}

In table \ref{nlsc} we display a summary of results for the SC model for
different values of the spectral index.
Note, to start with, that for large $n$ or more negative $\gamma$
the coefficients decrease with the perturbative order
(the subscript after the coma), indicating possible
convergence of the perturbative series. 
For $\gamma=0$ (\ie $n=-3$) the coefficients of each order
in the expansion increases quickly with the order, which might indicate
that the PT expansion does not converge for the unsmoothed field, 
at least in the SC approximation.

For $S_J$, there is a
suppression of non-linearities with the effects of
smoothing (as $n$ increases from $-3$ to $0$), which is
also found in numerical simulations (see \eg $\L$okas \etal 1995)). 
In particular,
vanishing non-linearities are found for $n \approx 0$.

\subsection{Comparison with Exact PT}
\label{sec:comp_ept}

The tree-level results in the SC model are of course identical to the ones
estimated in the exact PT by Juskiewicz \etal (1993) and B94b, as argued
in \S 2.2 . 
For the {\em unsmoothed} fields, analytic results including the 
first corrective term to the tree-level, were first derived by 
SF96a through loop calculations in the diagrammatic
approach to the exact PT. For the variance, skewness, and the
kurtosis (the last one only at tree-level) they obtain,
\bea
\sigma^2 \approx \sigma_l^2\,+\,1.82\,\sigma_l^4\,
+\,\Or \left(\sigma_l^6 \right) \nn \\
S_3 \approx 4.86\,+\,9.80\,\sigma_l^2\,
+\,\Or \left(\sigma_l^4 \right) \nn \\
S_4 \approx 45.89\,+\,\Or \left(\sigma_l^2 \right),
\label{exactpt}
\eea
for the average values in the range $2 \geq n \geq -2$ (for
a power-law power spectrum, $P(K) \sim k^n$), with a $3 \%$ 
variation within that range due to non-local effects.
The above results are to be compared to those from the SC in the 
perturbative regime (truncated at the 
same order),
\bea
\sigma^2 \approx \sigma_l^2\,+\,1.44\,\sigma_l^4\,
+\,\Or \left(\sigma_l^6 \right) \nn \\
S_3 \approx 4.86\,+\,10.08\,\sigma_l^2\,
+\,\Or \left(\sigma_l^4 \right) \nn \\
S_4 \approx 45.89\,+\,267.72\,\sigma_l^2\,
+\,\Or \left(\sigma_l^2 \right).
\eea
Comparing with the exact PT result Eq.[\ref{exactpt}], we can see
that tidal (non-local) 
effects only amount to a $3 \%$ in the corrective term 
for $S_3$ and up to $20 \%$ in
the first corrective term for the variance. The $S_4$ correction 
($\sim 268\,\sigma_l^2$) must be taken as an accurate 
prediction for $S_4$ (within the few per cent effect expected from 
the net tidal contribution for the $S_J$ ratios) 
since there are no analytic results available to 
compare with. We therefore see from the above comparison
that the shearless (SC) approximation leads to 
very accurate  predictions for
the hierarchical amplitudes, $S_J$,
while giving worse estimates for the cumulants, $\xi_J$.

We stress the importance of applying the SC 
approximation
in Lagrangian space, where is the exact spherically symmetric
solutions of the field equations.
The SC model is described by a transformation
that {\em only} depends on the value of the linear field at the same point
(what we call a {\em local-density transformation}). When
going back to Euler space the density fluctuation (defined at a point) 
is normalized with a factor
(see Eq.[\ref{normalag}]) which is a function of the (non-linear)
variance. Since the variance is
a volume average of the two-point correlation function, this factor 
yields some {\em non-local} contribution to the cumulants 
(in Euler space).
This {\em non-local} contribution
is missed when introducing the SC model in Euler space {\em directly},
thus is not surprising that the predictions for the 
cumulants in the SC approximation in Euler space are a poor estimation of those
in exact PT, as the latter are dominated by the
non-local (tidal) effects (see Table \ref{euler_tab} in
Appendix \ref{app:euler}). The contribution to the cumulants of 
this {\em non-local} term is typically negative so the predictions from the
SC model in Euler space generically overestimate those in Lagrangian space. 
However, the domination of non-local effects in the cumulants 
for the SC model 
in Euler space is partially canceled when computing the hierarchical
ratios $S_J$, similarly to what is found for the SC model in Lagrangian
space. We shall see below that {\em smoothing} effects do not alter
substantially this interpretation (at least for a top-hat window). 

For the {\em smoothed} fields, there are few analytic results which 
concentrate on the variance (and its Fourier transform, the power-spectrum), 
the skewness, and the 
bispectrum. For the present analysis we focus on the results for the 
variance and the skewness (see SF96b and Scoccimarro 1997, hereafter S97, 
respectively) which, in the 
diagrammatic approach to the exact PT to one-loop
order, for a top-hat smoothing and $n=-2$, give
\bea
\sigma^2 \approx \sigma_l^2\,+\,0.88\,\sigma_l^4\,
+\,\Or \left(\sigma_l^6 \right) \nn \\
S_3 \approx 3.86\,+\,3.18\,\sigma_l^2\,+\,\Or \left(\sigma_l^4 \right)   
\label{sfn2}
\eea
that we compare to the SC predictions,
\bea
\sigma^2 \approx \sigma_l^2\,+\,0.61\,\sigma_l^4\,
+\,\Or \left(\sigma_l^6 \right) \nn \\
S_3 \approx 3.86\,+\,3.21\,\sigma_l^2\,+\,\Or \left(\sigma_l^4 \right)   
\label{n=-2}
\eea
which means a $30\,\%$ and $1\,\%$ contribution from the tidal forces for 
the corrections in the
variance and the skewness, respectively.
These can be obtained by just 
subtracting the shearless (SC) contribution to the exact computation 
carried out by means of the loop calculations. 
Thus, we see that in line with the unsmoothed predictions, the
shearless contribution completely dominates the skewness $S_3$,
at least at the one loop order. 

\begin{figure}[t]
\centering
\centerline{\epsfysize=8.truecm 
\epsfbox{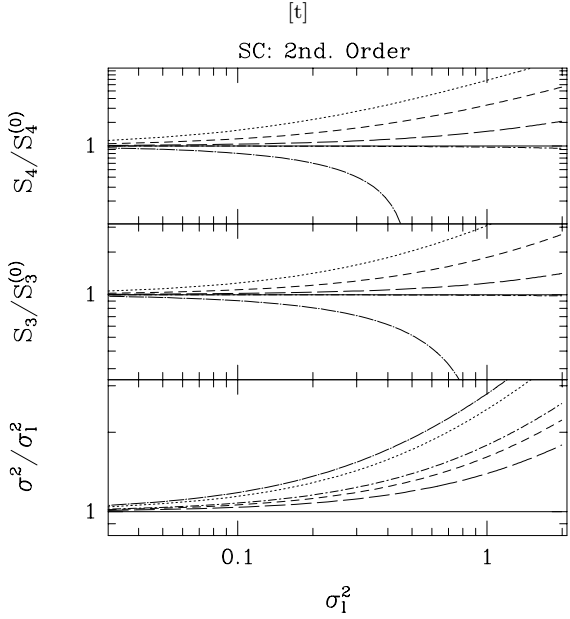}}
\caption[junk]{Departures from the tree-level 
contributions as the linear {\em rms} fluctuation grows, for the variance, 
skewness and kurtosis of the density field
as predicted by the SC model up to the 2nd non-vanishing 
perturbative  contribution (one-loop) for 
different values of the spectral index: the dotted line shows the $n=-3$
(unsmoothed) case, and the short-dashed ($n=-2$), long-dashed ($n=-1$),
dot short-dashed ($n=0$), dot long-dashed ($n=1$) depict the 
behavior for the smoothed density field. 
The solid line shows the tree-level values as a reference.
The corrective term has a minimum contribution to the variance for
$n \approx -1$, while the hierarchical amplitudes show a vanishing
one-loop contribution for $n \approx 0$.}
\label{scvpt1}
\end{figure}

\begin{figure}[t]
\centering
\centerline{\epsfysize=8.truecm 
\epsfbox{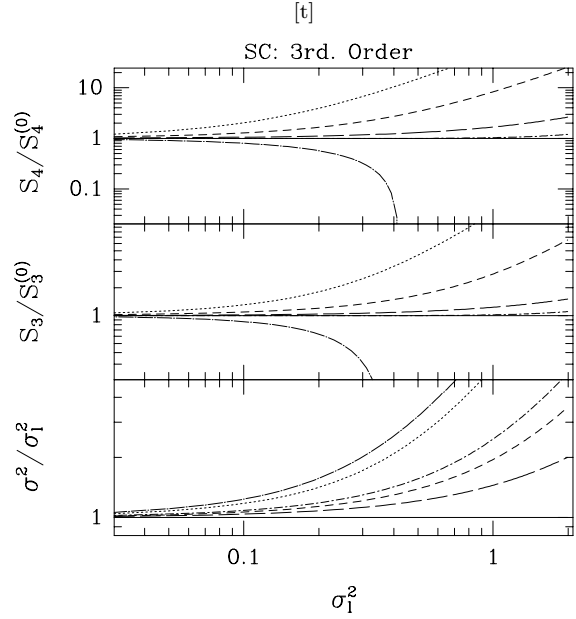}}
\caption[junk]{Same as Fig \ref{scvpt1} when the 3rd perturbative 
contribution (two-loop) is
included in the computation of the cumulants.}
\label{scvpt2}
\end{figure}

In Fig \ref{scvpt1} we display the departures from the tree-level contribution
for the evolved variance when the one-loop correction is included, 
skewness and kurtosis as the linear variance 
approaches unity, where the perturbative regime is expected to break down.
Fig \ref{scvpt2} shows the same as Fig \ref{scvpt1} when 
the 3rd  order (3rd non-vanishing
contribution) is included in the computation of the cumulants: the 
qualitative features are preserved when this higher-order contribution is
taken into account with respect to those inferred from the 2nd order analysis.
Only a monotonic enhancement of the scaling properties is observed.

In Table \ref{comp} we compare the results from different approximations 
to the exact dynamics in the perturbative regime (exact PT), available in the 
literature and we give as well   
the SC predictions in the same regime. We recall that the FFA is based
on a linearization of the peculiar velocity field. The LPA assumes the 
potential remains linear throughout the gravitational evolution, and the ZA
takes the trajectories of the particles to be straight lines. 
All values concerning the one-loop contribution 
(including the exact PT ones), except for the SC ones, 
are derived 
making use of the diagrammatic formalism and are given in SF96a. Tree-level
amplitudes had been derived previously though and a summary of those 
results may be found in Bernardeau \etal (1994). 
As summarized in Table 3, the SC model yields the best
estimates for the loop corrections in PT within the available
non-linear approximations.

\begin{table}

\begin{center}

\begin{tabular}{|c||c|c|c|c|c|}
\hline \hline
Dynamics & $s_{2,4}$ & $S_{3,0}$ & $S_{3,2}$ & $S_{4,0}$ & 
$S_{4,2}$\\ 
\hline \hline
$FFA^{*}$  & 0.43 & 3 & 1 & 16 & 15.0 \\ 
\hline
$LPA^{*}$   & 0.72 & 3.40 & 2.12 & 21.22 & 37.12 \\
\hline
$ZA^{*}$ & 1.27 & 4 & 4.69 & 30.22 & 98.51 \\
\hline
$SC$ & 1.44 & 4.86 & 10.08 & 45.89 & 267.72 \\ 
\hline \hline
$Exact\ PT^{*}$ & 1.82 & 4.86 & 9.80 & 45.89 & $?$ \\ 
\hline \hline 
\end{tabular}

\caption[junk]{Comparison between different non-linear approximations to the 
for the {\em unsmoothed} fields up to the first corrective term beyond 
tree-level (the {\em one-loop} term). 
The asterisk denotes the results obtained within the diagrammatic
approach for the relevant dynamics.}
\label{comp}

\end{center}

\end{table}

\subsection{Previrialization in the Variance}
\label{sec:virial}

It is important to notice the different behavior between 
the cumulants (like
the variance) and the hierarchical amplitudes, $S_J$. 
It has long been argued that non-local evolutionary
effects generate a suppression of collapse on the largest scales due to the
increase in small-scale power that generates large scale random motions, the  
so-called effect of {\em previrialization}
(see Davis and Peebles 1977, Peebles 1990, $\L$okas \etal 1996). 
This will result in a non-linear variance that is smaller
than the linear one within a certain range. 
This effect has been found in N-body
simulations  (\eg see Figure 9 in BGE95).

\begin{figure} 
\centering
\centerline{\epsfysize=8.truecm 
\epsfbox{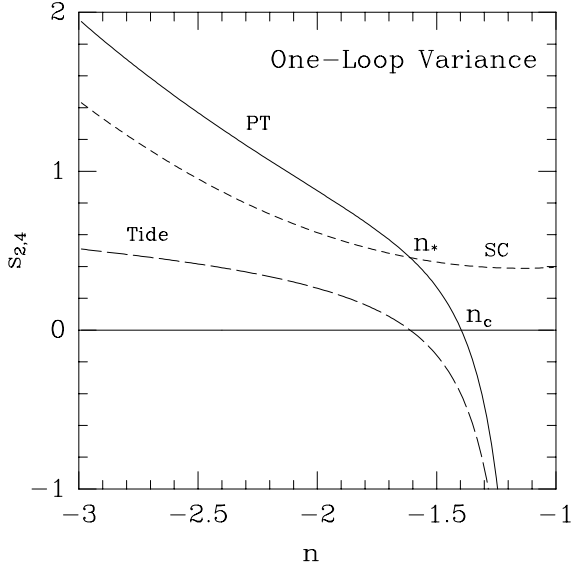}}
\caption[junk]{
The one-loop contribution to the variance in PT.
The solid line depicts the exact PT behavior, the short-dashed
line shows the shearless (SC) contribution while the long-dashed line
gives the tidal contribution. Two characteritic indices arise in this
analysis: $n_c \simeq -1.4$ where the one-loop variance vanishes, and
$n_{\star} \simeq -1.6$, where tidal effects are zero and the
SC model prediction is exact.}  
\label{loop_var}
\end{figure}

The local picture arising from the SC model is unable 
to account 
for a complete suppression of non-linearities 
in the large scale density (and velocity)
fields and yields a faster growth of fluctuations on the large  scales
compared to the actual non-local dynamics.
This qualitative feature of the SC model was already pointed out
by Bernardeau (1994c), 
in what he called the {\em rare event limit}, 
$\delta_l/\sigma_l \rightarrow \infty$. Making use of the
third order in PT (in Lagrangian space) for the density field
{\em smoothed} with a top-hat window (see his Eq.[38]) for a scale free
power spectrum, 
he claimed that the SC
picture deviates from the exact PT
as smoothing effects increase.
In particular, he found that for a spectral index $n \simgt -1$, 
PT diverges due
to the shear effects. Whenever more realistic spectra 
(models with a cut-off on small scales, such as CDM or APM-like ones) are 
introduced, this divergence becomes only logarithmic.
 
In PT, the above commented
suppression depends on the smoothing effects and leads to
the appearance of a critical value of the spectral index (assuming
scale free power spectrum) around $n_c \approx -1.4$ (see S97), for which 
non-linear contributions vanish. Figs
\ref{scvpt1} and \ref{scvpt2} illustrate well this point since no 
change of sign in the
first non-linear contribution (beyond tree-level) to the evolved
variance is appreciated in the plot contrary to what has been 
found in numerical simulations depicting the 
exact perturbative scaling (see Lokas \etal 1996, Fig 4). 
Note that the SC prediction yields a minimum 
contribution at $n \simeq -1.14$, which is not far from 
$n_c \approx -1.4$, but the net effect at the minimum is
non-zero.

 Comparing the results for the variance from the exact PT and the shearless
(SC) approximation, we can infer some information about where the tidal 
effects become lower, comparable or greater than the shearless contribution, 
provided the variance changes smoothly. Let us 
formally decompose the one-loop contribution
to the variance in the shearless (SC) and tidal contributions as follows,
\beq
s_{2,4}^{PT} = s_{2,4}^{SC} +s_{2,4}^{Tide} ,
\eeq
where $s_{2,4}^{SC} > 0$ for all $n$ as displayed on the bottom panel of 
Fig \ref{scvpt1}. 
The one-loop contribution to the variance in PT, $s_{2,4}^{PT}$ 
can be integrated from the analytic expression for the one-loop
power-law power spectrum given in SF96b (see their Eq.[B5]) 
valid in the range $-1>n>-3$.
As shown in Fig. \ref{loop_var}, the shearless (SC) contribution
gives a systematic underestimation of the PT value from the unsmoothed
value ($n=-3$) up to $n_{\star} \simeq -1.6$ where the tidal contribution
vanishes, and thus, $s_{2,4}^{PT} = s_{2,4}^{SC}$. Around this index, 
the SC model gives an accurate estimation of the one-loop variance.
Beyond this point, $n \simgt n_{\star}$, the tidal contribution 
rapidly grows and becomes negative (see Fig.\ref{loop_var}). 
For $n_c \simeq -1.4$, $s_{2,4}^{SC}= -s_{2,4}^{Tide}$ 
and the one-loop variance vanishes. As seen in the plot, while the
shearless contribution remains positive and small, the tidal contribution
takes larger and larger negative values as $n \rightarrow -1$, what
breaks down the PT approach (at least for the variance). 

\begin{figure} 
\centering
\centerline{\epsfysize=8.truecm 
\epsfbox{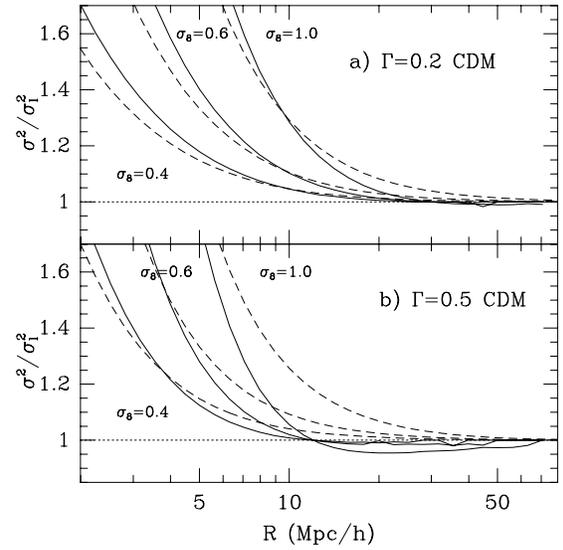}}
\caption[junk]{The variance
in spherical shells of radius $R$. We compare linear theory
to the SC model and the exact PT
(leading order) non-linear correction, against the linear term. 
Each panels, (a) and (b), 
show the ratio of the non-linear to the linear
variance for the  $\Gamma=0.2$ (a) or $\Gamma=0.5$ (b)
CDM initial $P(k)$.
In both cases the continuous line from bottom to top
corresponds to $\sigma_8=0.4, 0.6, 1.0$. 
The SC predictions for the same amplitudes are shown as 
dashed lines.}
\label{sigpt}
\end{figure}

In Figure \ref{sigpt} we compare the 
non-linear evolution of the variance for the SC to that of the exact PT.
We present the ratio of the non-linear to the linear
variance from CDM model, 
with $\Gamma=0.2$ (top panel) and $\Gamma=0.5$ (bottom),
as a function of the smoothing radius.
The leading order non-linear SC predictions are shown 
as dashed lines. The corresponding (numerically integrated) PT 
results are shown as continuous lines.
These are estimated from $P(k)$ to second
order (2nd non-vanishing contribution) 
as obtained following Baugh \& Efstathiou (1994).
In both cases we show results for different amplitudes
of the linear variance: $\sigma_8=0.4, 0.6, 1.0$,
which has a linear variance $\sigma \simeq 0.5$ at
scales $R \simeq 6, 10, 15 \Mpc$, respectively.
We can define an {\em effective index}, $n_{eff}$, as the index 
at the scale where the variance becomes unity. For 
the $\Gamma=0.2$ model we have that $n_{eff}^{\Gamma=0.2} \simeq -2$,
while for $\Gamma=0.5$ this is about $0.5$
larger (this can be seen for example in Figure 9 in Croft
\& Gazta\~naga 1998):
 $n_{eff}^{\Gamma=0.5} \simeq -1.5$.
The SC prediction should match
better the  $\Gamma=0.2$ model, as the effective index is closer
to $n_{\star}$, where tidal effects vanish. 
For the $\Gamma=0.5$ model we have that
$n_{eff} \simeq  n_c$ and the SC model is not such a good approximation. 
In this
case we expect non-linear effects to be small due to the cancellation
of tidal and purely local contributions to the cumulants. 
This can be seen in Figure \ref{sigpt}.
The SC model matches well the PT prediction for the $\Gamma=0.2$ model,
where non-linear effects are more important.  It also matches well
the earlier evolution of $\Gamma=0.5$ model, for small amplitudes 
of $\sigma_l$, \eg earlier outputs
given by smaller $\sigma_8$. 
For the later outputs (\eg $\sigma_8=1$ in Figure \ref{sigpt}, bottom panel)
tidal effects
become important canceling out the non-linear growth (previrialization
effect).
This is not fully recovered by the SC model (which only accounts for
the local effects). Nevertheless note that
for small scales, where the variance is large,  the prediction
seems to be dominated by the local effects and the SC model also
recovers the exact prediction.

\begin{figure} 
\centering
\centerline{\epsfysize=8.truecm 
\epsfbox{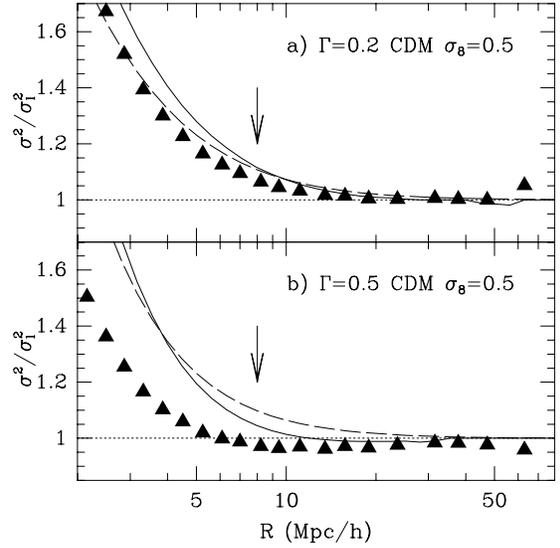}}
\caption[junk]{Non-linear evolution of the variance.
Symbols show the ratio of the non-linear to the linear
variance from CDM N-body simulations
with $\Gamma=0.2$ (top panel) and $\Gamma=0.5$
(bottom panel) as a function of the smoothing radius.
In both cases $\sigma_8=0.5$.
The SC model predictions are shown as 
a short-dashed line while PT predictions are shown
as a continuous line. The arrows indicate where $\sigma_l=0.5$.}
\label{sig05}
\end{figure} 

\begin{figure} 
\centering
\centerline{\epsfysize=8.truecm 
\epsfbox{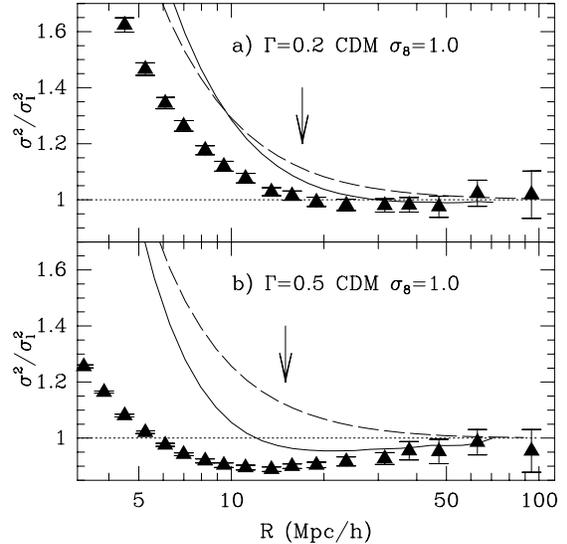}}
\caption[junk]{Same as Fig. \ref{sig05} for
$\sigma_8=1.0$.}
\label{sig10}
\end{figure} 

In Figure \ref{sig05} we show the 
non-linear evolution of the variance
as the ratio to  the linear
variance from CDM N-body simulations
with $\Gamma=0.5$ (filled triangles) and $\Gamma=0.2$
(open triangles) as a function of the smoothing radius.
The error-bars, from 10 realizations of the models, which are
not displayed for clarity, are always smaller
than the symbols except for the last $3$ points where the
error-bars  are smaller than twice the size of the symbols. 
In both cases $\sigma_8=0.5$ and 
the linear variance $\sigma \simeq 1$ at $R \simeq 2 \Mpc$.
The SC predictions are shown as dashed lines, which are to
be compared to the exact
PT results (continuous lines). For the  $\Gamma=0.2$ model
there is hardly any difference between the predictions and
the N-body results. In fact, the latter follows both predictions up to large
values of $\sigma$ (small scales). For the
$\Gamma=0.5$ model, non-linear effects are small during the
weakly non-linear phase and both linear and non-linear
predictions are close to the N-body results. There is a
slight sign of previrialization at $R \simeq 20 \Mpc$, but the
effect is quite small. This way, despite the exact PT results are
more accurate in qualitative terms, there is little
difference in practice. Note that, in this case, the agreement
with predictions does not extend to large values of $\sigma$ 
(small scales) unlike the case of the $\Gamma=0.2$ model.

\begin{figure} 
\centering
\centerline{\epsfysize=8.truecm 
\epsfbox{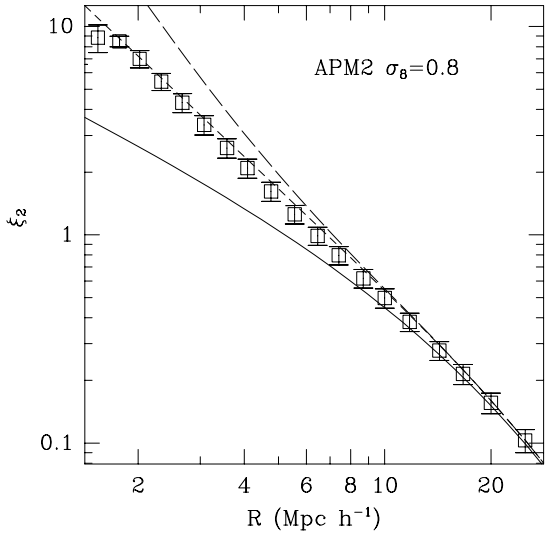}}
\caption[junk]{The variance
for APMPK
at $\sigma_8=0.8$ (figures) 
 compared to PT models (lines):
(a) linear theory
prediction (continuous line),
 (b) the 2nd order result, \ie when the 2nd perturbative contribution 
to the variance is included 
in the SC (short dashed lines) 
 (c) the 3d order result  in the SC (long dashed lines).}
\label{x2apmnl}
\end{figure}

Figure \ref{sig10} shows the corresponding results for
$\sigma_8=1$. In this case, the $\Gamma=0.2$ CDM model
is the one with an effective index $n_{eff} \simeq n_c$
so that non-linear effects are small
and we have a similar situation to the $\Gamma=0.5$ model
at $\sigma_8=0.5$. The effective index for 
$\Gamma=0.5$ model at $\sigma_8=1.0$ si $n> n_c$ so that the
previrialization effect discussed above is larger and the SC
prediction fails (long-dashed line).
The prediction for the second order contribution to $P(k)$ for  
$\Gamma=0.5$ is shown as a continuous line. 
Although it shows
some previrialization effect (the ratio is smaller than unity),
it only fits the N-body results in the narrow range of scales,
$R \simgt 30 \Mpc$, where linear theory is a good 
approximation, given the errors.
The above arguments explain
in a simple fashion why the SC model matches so well the variance of the
APM simulations even on small scales (for
output times $\sigma_8 \simlt 1$)
as shown in Fig. \ref{x2apmnl}. The effective index 
$n_{eff}$ on the quasi-linear 
scales is within the range $-2 \simgt n_{eff} \simgt -1.5$, a bit lower
than the $\Gamma=0.2$ CDM model (see Fig.3 in BG96, where
quasi-linear scales correspond to $k \simeq 0.1$). For this index,
tidal effects  are expected to give a sub-dominant contribution. 
Although 
the SC model seems to match well the simulations up to the scales where the
linear variance $\sigma_l \simgt 1$, \eg beyond the point where PT
must break down,  the third order SC correction (long-dashed line)
shows significant deviations already 
at $R\simeq 6 \Mpc$, where $\sigma_l \simeq 1$.

The results here obtained are in qualitative agreement with those found
by B94c, according to which, the {\em rare event limit}
progressively differs from the exact PT predictions as 
$n \rightarrow -1$, even 
in the first correction to the variance $s_{2,4}$. 
Furthermore, B94c also finds no indication of {\em previrialization}
in the {\em rare event limit} and stresses the unimportance of
this effect for the APM \& IRAS observations, given the 
effective spectral index related to the scales on
which they lie ($n_{eff} \simeq -1.3, -1.4$, respectively).
However, according to B94c, the {\em rare event limit} is {\em always}
different from the exact PT result whenever $\delta_l \neq 0$. This
is at variance with what we find here, where the SC model is expected
to yield accurate results, particularly for values of the spectral index
$n \in [-1.5,-2]$. What is more, there must be some value $n_{\star}$ 
within that interval for which tidal effects exactly vanish.

\subsection{Comparison of the $S_J$ Ratios with N-body Simulations: 
Previrialization Lost}

On the other hand, when it comes to evaluating the hierarchical 
amplitudes, $S_J$,
it turns out that the cancellation of (non-local) tidal effects 
erases the
previrialization effect and the suppression of non-linearities at the first
corrective order beyond tree-level appears
at a different spectral index $n \approx 0$, what seems to be
an intrinsic (local) shearless feature if the good agreement between 
the local (SC) and
non-local (exact PT) hierarchical amplitudes is anything to go by.
Furthermore, loop-calculations in Fourier space concerning the 
reduced Bispectrum, $Q$, seem to confirm this argument as
the scale dependence of $Q$ retains non-vanishing 
non-linearities
even at $n \approx -1.5$ (at least for the equilateral configuration, see
Scoccimarro \etal 1998, Eq.[30]).
However, the reduced bispectrum also preserves some 
non-local features which manifest  
in terms of a loss
of configuration dependence (isotropization) roughly at the same value 
of the spectral index as the one estimated from the variance or the power
spectrum, $n_c \approx -1.4$ (see SF96b, and 
S97). 


\begin{figure} 
\centering
\centerline{\epsfysize=8.truecm 
\epsfbox{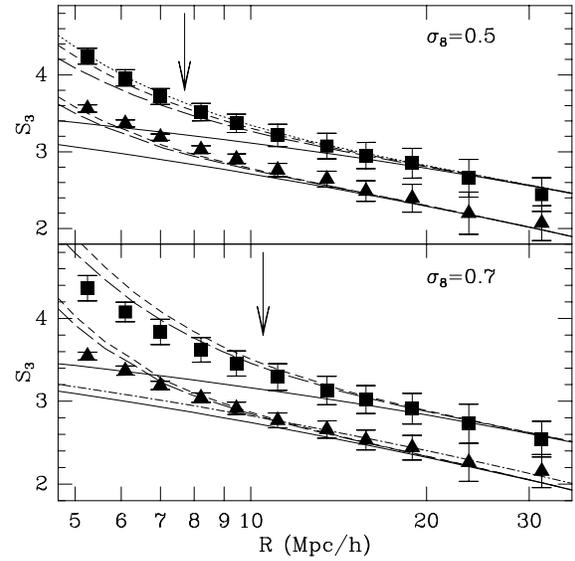}}
\caption[junk]{The hierarchical skewness $S_3$
from 10 realizations of flat CDM N-body simulations
at output time $\sigma_8=0.5$ (top panel) and 
$\sigma_8=0.7$ (bottom panel). Symbols 
with error-bars correspond to $\Gamma=0.2$ (squares), 
and  $\Gamma=0.5$ (triangles).
Each case is compared to the corresponding PT predictions (lines):
a) the PT tree-level
prediction uncorrected for the ZA transients (dot-dashed line), 
and corrected for
the transients (continuous), b) the 2nd order result  
in the SC  uncorrected for the ZA transients (short-dashed), 
corrected for the ZA transients (long-dashed), and the case with
$\gamma_2 = 0$ (dotted).
The arrows show where $\sigma = 0.5$.
The dot-dashed line shows the tree-level prediction for 
the  $\Gamma=0.5$ and $\sigma_8=0.7$, uncorrected for the ZA transient.}
\label{s3cdm}
\end{figure}

In Figs \ref{s3cdm}, we compare the SC predictions with  
CDM simulations for two different models. The smoothing
coefficients $\overline{\nu_k}$ are modified to include
the higher-order logarithmic derivatives of the variance 
(see Eq. [\ref{tophatgm}] in Appendix \ref{app:locumul}) 
but it turns out that, within the errors,
they do not significantly
change the predictions derived from the power-law power spectrum.
At small scales $\gamma_2$ changes $S_3$ by less than $10\%$, which hardly
makes any difference given the errors, as shown by the 
dotted line, only displayed for the $\Gamma=0.2$ $\sigma_8=0.05$ model
for clarity.
The error bars shown in the SCDM plot are derived from $10$ realizations.  
The arrows in the Figure show the scale at which $\sigma=0.5$.

On large scales the spectral index $n$ for CDM models
becomes larger 
and non-linearities are very small, as predicted by the SC model
(see Table \ref{nlsc}). This explains why deviations from the
tree-level are hardly noticeable in the later outputs (bottom
panel) and they only become important, given the errors, 
when PT theory should break down
($\sigma_l \simeq 1$).
The effective index in the earlier outputs is smaller (corresponding
to smaller scales) and deviations from tree-level are more significant.
In this regime the SC predictions show a very good agreement with
the N-body simulations (top panel) up to $\sigma \simeq 1$.
Note that the predictions for the exact PT corrections are not available
in this case, but the agreement with the N-body simulations indicates that
the SC models provides an excellent approximation.


We have corrected for ZA transients, 
an artifact due to the initial (ZA) start in the N-body
simulations (see BGE95 1995, Scoccimarro 1998). 
We use the analytic results of Scoccimarro (1998) for the
tree-level:
\beq
S_{3,0} = {34\over{7}} + \gamma -{2\over{5}} a^{-1}  -{16\over{35}} a^{-7/2}
\eeq
where $a$ is the expansion factor away from the (ZA) IC.
In principle 
this is not a very important correction for the tree-level result, 
as the sampling errors on large scales, where $\sigma$ is small, are large. 
This is more important for smaller scales, 
as the errors are smaller. In 
Figs \ref{s3cdm} we show the tree-level prediction for 
$a \rightarrow \infty$ (dot-dashed line) as compared to the
actual prediction, including the transient, for $a= 5.0$ (the bottom 
continuous line) in the $\Gamma=0.5$ model for $\sigma_8=0.7$. 
The long-dashed line shows the 2nd order (one-loop) prediction 
in the SC model corrected for the ZA transients. This
can be done by using the tree-level results for the transients, as they 
determine all higher-orders in the SC model.
As pointed out by Scoccimarro, the effects of the transients tend  
to create the false impression that the tree-level PT has a wider
range of validity. In our case, by comparing the points to the 
dot-dashed line one might conclude that tree-level results are valid up to
$R \simeq 9 \Mpc$, while a comparison to the corrected prediction (continuous
line) indicates $R \simeq 15 \Mpc$. On the other hand, adding the loop 
terms to the uncorrected tree-level would give the  false impression 
of a poor agreement with the SC model.



\begin{figure} 
\centering
\centerline{\epsfysize=8.5truecm 
\epsfbox{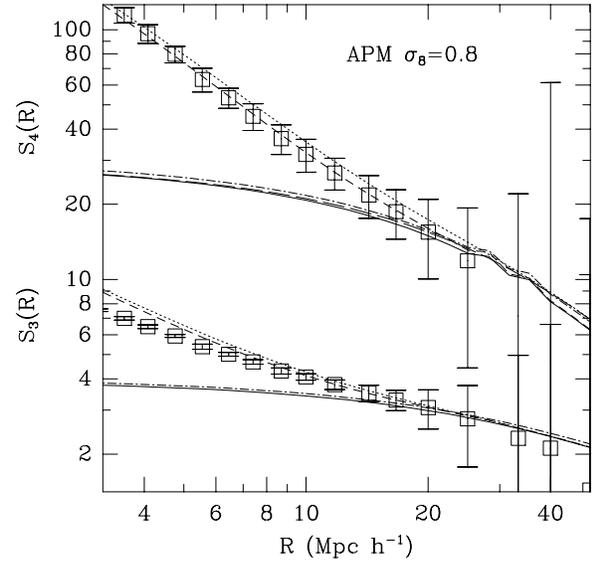}}
\caption[junk]{The skewness, $S_3$, and kurtosis, $S_4$
for simulations of the APMPK 
at $\sigma_8=0.8$ (open squares with error-bars) 
 compared to  tree-level  theory
prediction (continuous line) and
the 2nd order result ($S_J$ including the one-loop term) 
in the SC (short dashed lines).
The corresponding predictions for $\gamma_2=0$ are shown as
long-dashed and dotted lines respectively. The dot-dashed lines
show the tree-level predictions uncorrected for the ZA transients.
}
\label{s34apmnl}
\end{figure}
\begin{figure} 
\centering
\centerline{\epsfysize=8.5truecm 
\epsfbox{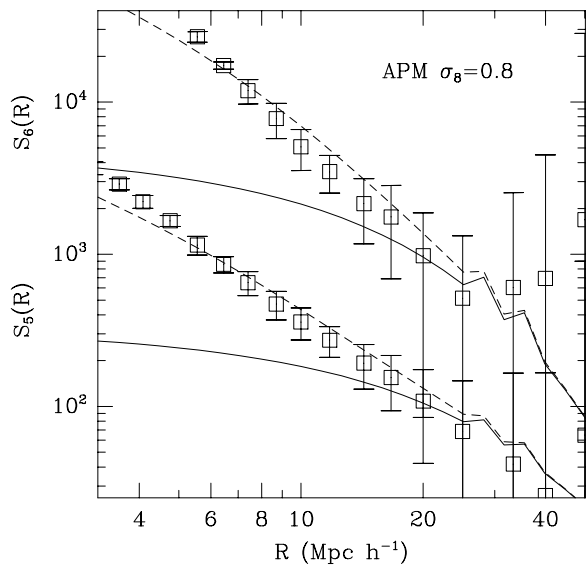}}
\caption[junk]{Same as Fig. \ref{s34apmnl} for $S_5$ and $S_6$.}
\label{s56apmnl}
\end{figure}

Figs \ref{s34apmnl} and \ref{s56apmnl} display the scaling of the higher-order
moments according to the APM-like simulations (the mean of APMPK1 and
APMPK2 in Table \ref{Nbody}). As for the previous cases, the
SC model matches well the observed behavior for the higher-orders in the 
N-body 
simulations up to the scales where PT is expected to break down, \eg 
$\sigma_l \approx 1$. 
As we consider higher order moments, deviations from the tree-level
prediction are larger, a trend that is reproduced by the SC predictions.
We have found a similar trend for higher orders in the CDM models.

The ZA transients, shown as dot-dashed lines in Fig. \ref{s34apmnl},
are not very important in this case ($a\simeq 6$). The effect of 
using $\gamma_2=0$ is also small although it tend to introduce a slight
shift in the predictions. 

These results extend the domain of applicability of the SC model in PT
beyond
the point expected from B94c analysis of the {\em rare event limit},
as potential cancellations of the tidal contributions in the $S_J$
ratios were not examined there.

\section{Discussion and Conclusions}
\label{discuss}

No available analytic approximations to the dynamics of 
cosmological perturbations lead to accurate predictions
for the one-point cumulants in the non-linear regime (see \S\ref{sec:other}).
This applies to all popular approximations, such as those based on putting some
constraints on any of the fields that couple to the density 
(such as the FFA or the LPA), the
Zel'dovich Approximation (ZA), which puts a constraint
on the trajectories of the masses, or the local 
Lagrangian approximations, which
assume that the evolution of the density contrast may be described locally,
\ie without any influence of the surrounding matter to the
 mass trajectory. Despite giving some useful insight to the exact dynamical
picture with much simpler calculations, none of them is able to  
reproduce the exact values of the one-point cumulants even at tree-level as 
derived in the exact PT.

As mentioned in the introduction, there are serious difficulties in 
applying the exact PT approach to find the gravitational evolution 
of the cumulants 
of a Gaussian field  to higher-order (loop) corrections, \ie
to compute next-to-leading orders. 
For non-Gaussian initial conditions this problem is apparent even to leading 
order (in non-linear corrections). 
An important simplification is found when
only the {\it monopole} or spherical contribution is considered.
This contribution is exact for all the tree-graphs.
We have shown  how the solution in this case 
can be found directly from the dynamical
equations of the evolution of a cosmic
fluctuating field (see \eg Appendix \ref{sec:tree}
and Appendix \ref{app:za}) as it is given by the spherical collapse (SC) model
for the case of density fluctuations 
for a non-relativistic presureless  irrotational fluid. 
This provides us with a
simpler derivation and interpretation of the results presented
by Bernardeau (1992), showing 
why the SC model gives the exact tree-level in PT. This lacked a
satisfactory explanation in the work of Bernardeau.

We have explored the predictions for the
one-point cumulants of the density field 
in the weakly-nonlinear regime within the {\em spherically
symmetric} approximation to the dynamics.
This is done both for the exact dynamics 
and for the Zel'dovich approximation (see Appendix \ref{app:za}).
As the SC model gives the exact tree-level irrespective of the nature of
the initial fluctuations and the value of $\Omega$, it 
can also be used to make predictions for non-Gaussian initial
conditions (the results are presented in the accompanying Paper II), 
the velocity field and 
the $\Omega$-dependence of these predictions 
(see Paper III of the series).

A natural shortcoming of 
the SC model is the loss of {\em previrialization}, \ie the
generation of 
large scale random motions induced by the small scale power, which is a
non-local phenomenon. 
This is in qualitative agreement with the analysis 
given in B94c in the {\em rare event limit} to PT which 
is described by the SC model.
The lack of {\em previrialization} is
illustrated by the absence of vanishing non-linearities in the variance
for a critical index $n_c \approx -1.4$ (see \S\ref{sec:virial}),
which is a direct consequence of  {\em previrialization}.
This phenomenon is 
expected from the exact PT (SF96b) and it has been observed
in numerical simulations ($\L$okas \etal 1995).

Despite this, we have shown that the SC model also yields accurate
predictions for the cumulants beyond the tree-level.
Tidal effects in the variance are actually small because the 
effective index $n_{eff}$,
as measured in galaxy catalogues (such as the APM),
 is $n_{eff} \simeq [-1, -2]$
(see \S\ref{sec:virial}) for which either non-linearities 
are small or tidal
effects are sub-dominant.
 
We stress the importance of applying the SC 
approximation
in Lagrangian space. There, the SC model is described by a transformation
that {\em only} depends on the value of the linear field at the same point
(what we call a {\em local-density transformation}). However, when
going back to Euler space the density fluctuation (defined at a point) 
is normalized with a factor
(see Eq.[\ref{normalag}]) which is a function of the (non-linear)
variance. Since the variance is
a volume average of the two-point correlation function, this factor 
yields some {\em non-local} contribution to the cumulants 
(in Euler space).
This {\em non-local} contribution
is missed when introducing the SC model in Euler space {\em directly},
thus is not surprising that the predictions for the 
cumulants in the SC approximation in Euler space are a poor estimation of those
in exact PT, as the latter are dominated by the
non-local (tidal) effects (see Table \ref{euler_tab} in
Appendix \ref{app:euler}). The contribution to the cumulants of 
this {\em non-local} term is typically negative so the predictions from the
SC model in Euler space generically overestimate those in Lagrangian space. 

For the predictions within the SC model (in Lagrangian space) 
for the hierarchical ratios, $S_J$,
tidal effects partially cancel out, at least to one-loop. 
We find an excellent agreement for the hierarchical amplitudes $S_J$
in the perturbative regime of the SC model, with those
derived by SF96a for the exact PT in the diagrammatic approach.
A similar conclusion follows for the Zel'dovich dynamics 
(Appendix \ref{app:za}),
where the monopole approximation (the SSZA) results  
are in  excellent agreement with the exact calculations.

We have also compared the predictions for the higher-order
moments from the SC model to those measured in 
CDM and APM-like N-body simulations,
and they turned out to be in  
very good agreement in all cases up
to the scales where $\sigma_l \approx 1$, supporting our view that
the tidal effects have only a marginal contribution to the reduced
cumulants. Furthermore, the
break down of the shearless approximation roughly coincides with the regime
for which the perturbative approach itself breaks down, 
$\sigma \simeq 0.5$. That is,
where the contribution of the second (one loop) and the third 
(two loop) perturbative order in the SC model are significantly different.

These results extend the domain of applicability of the SC model in PT
beyond the point expected from B94c analysis of the {\em rare event limit},
as potential cancellations of the tidal contributions in the $S_J$
ratios were not examined there.

Future work intended to extend the SC model 
should try to incorporate the 
local contribution to the tidal field by including other
multipoles in the kernels $F_n$, \eg including the shear 
by using the dipole contribution.

\section*{Acknowledgments}

We want to thank Roman Scoccimarro, Francis Bernardeau
and Josh Frieman 
for carefully reading the 
manuscript and pointing out useful remarks. 
EG acknowledges support from CIRIT (Generalitat de
Catalunya) grant 1996BEAI300192. 
PF acknowledges a PhD grant supported by 
CSIC, DGICYT (Spain), projects PB93-0035 and PB96-0925.
This work has been
supported by CSIC, DGICYT (Spain), projects
PB93-0035, PB96-0925, and CIRIT, grant GR94-8001.

\section{References}

\def\refe {\par \hangindent=.7cm \hangafter=1 \noindent}
\def\aj {ApJ, }
\def\aa {A \& A, }
\def\prl {Phys.Rev.Lett., }
\def\ajs{ApJS, }
\def\mn {MNRAS, }
\def\mnl {MNRAS Lett., }
\def\apl {ApJ.Lett., }

\refe Bagla, J.S., Padmanabhan, T., 1994, \mn  266, 227
\refe Bardeen, J., Bond, J.R., Kaiser, N., Szalay, A.S., 1986 \aj 304, 15
\refe Baugh, C.M., Efstathiou, G., 1994, \mn 270, 183
\refe Baugh, C.M., Gazta\~{n}aga, E., Efstathiou, G., 1995, \mn 274,
1049 (BGE95)
\refe Bernardeau, F., 1992, \aj 392, 1 (B92) 
\refe Bernardeau, F., 1994a, A\&A 291, 697 (B94a) 
\refe Bernardeau, F., 1994b, \aj 433, 1 (B94b)
\refe Bernardeau, F., 1994c, \aj 427, 51 (B94c) 
\refe Bernardeau, F. \& Kofman, L., 1995 \aj 443, 479 
\refe Bernardeau, F., Singh, T.P., Banerjee. B., Chitre, S.M., 1994, 
\mn 269, 947 
\refe Bertschinger, E., Jain, B., 1994., \aj 431, 486
\refe Brainerd, T., Scherrer, R.J., Villumsen J. V., 1993, \aj 418, 570
\refe Colombi, S., Bouchet, F.R., Hernquist, L., 1996, \aj 465, 14
\refe Croft, R.A.C., Gazta\~{n}aga, E., 1998, \aj 495, 554
\refe Davis, M., Peebles, P.J.E., 1977, \ajs 34, 425
\refe Fosalba, P. \& Gazta\~{n}aga, E., 1998, \mn in press 
(Paper III, this issue)
\refe Fry, J.N., 1984, \aj 279, 499
\refe Fry, J.N., Gazta\~naga, E., 1993, \aj 413, 447 
\refe Gazta\~naga, E. \&  Baugh, C.M., 1995, \mnl 273, L1
\refe Gazta\~naga, E. \&  Fosalba, P., 1998, \mn in press 
(Paper II, this issue)
\refe Goroff, M.H., Grinstein, B., Rey, S.J., Wise, M.B., 1986, \aj 311, 6
\refe Hui, L., Bertschinger, E., 1996, \aj 471, 1  
\refe Jain, B., Bertschinger, E., 1994., \aj 431, 495
\refe Juszkiewicz, R., Bouchet, F.R., Colombi, S., 1993 \apl 412, L9
\refe Kaiser, N., 1984, \apl 284, L9
\refe Kofman, L., Pogosyan, D., 1995, \aj 442, 30
\refe \L okas, E.L., Juszkiewicz, R., Weinberg, D., Bouchet, F.R., 1995, 
\mn 274, 730 
\refe \L okas, E.L., Juszkiewicz, R., Bouchet, F.R., Hivon, E., 
1996, \aj 467, 1
\refe Matarrese S., Lucchin F., Moscardini L., Saez D., 1992 \mn 259, 437  
\refe Matsubara, T., 1994, \aj 434, 43
\refe Mo, H.J., White, S.D.M., 1996, \mn 282, 347
\refe Munshi, D., Sahni, V., Starobinsky, A.A., 1995,
Astron. Soc. of India Bulletin, 23, 564 
\refe Peebles, P.J.E., 1980, {\it The Large Scale Structure of the 
Universe:} Princeton University Press, Princeton
\refe Peebles, P.J.E., 1990, \aj 365, 27
\refe Press, W.H., Schechter, P., \aj 187, 425
\refe Protogeros, Z.A.M.,Scherrer,R.J., 1997, \mn 284,425 (PS97)
\refe Scoccimarro, R., 1997, \aj  487, 1 (S97) 
\refe Scoccimarro, R., 1998, \aj submitted, preprint astro-ph/9711187 
\refe Scoccimarro, R., Frieman, J., 1996a, \ajs 105, 37 (SF96a) 
\refe Scoccimarro, R., Frieman, J., 1996b, \aj 473, 620 (SF96b)
\refe Scoccimarro, R., Colombi, S., Fry, J.N., Frieman, J., Hivon, E.,
Melott, A., 1998, 496, 586 
\refe Zel'dovich, Ya.B., 1970. \aa 5, 84


\appendix

\section[Cumulants in Lagrangian Space]
{Cumulants from the Local-Density Transformation of Gaussian Initial 
Conditions in Lagrangian Space}
\label{app:locumul}

For a generic model of non-linear evolution of the density field in which
all the information is encoded in the linear density field alone, one can
construct a transformation of the kind,
 \beq
\delta=\La[\delta_1]=\sum_{n=1}^{\infty}\,{c_n \over n!}\,[\delta_l]^n
\eeq 
(see Eq.[\ref{local}]) with $c_1=1$, to reproduce the linear solution. 
This is what we shall call the {\em local-density} transformation. 
If we
further assume that the above transformation describes the 
non-linear evolution in Lagrangian space, it has to be normalized

Next we present the general results for the cumulants including their
first non-vanishing perturbative orders for GIC. 
For the SC model the calculation corresponds
to the process described in \S\ref{sec:sc}.
Note that this can be extended 
in a straightforward way to NGIC, just by taking into account the relevant 
terms to be kept in the perturbation series (see Paper II).

We proceed and give the coefficients of the perturbative terms according to
the notation introduced in \S\ref{sec:comparison},   
\bea
s_{2,4} &=& 3 - 4 c_2  + c_2^2 /2 + c_3 \nn \\   
s_{2,6} &=& 15 - 36 c_2  + 81 c_2^2 /4 - 3 c_2^3 /2 + 10 c_3  - 7 c_2  c_3 \nn \\  
    &+& 5 c_3^2 /12 - 7 c_4 /4 + c_2  c_4 /2 + c_5 /4  \nn \\ 
S_{3,0} &=& 3 c_2  \nn \\ 
S_{3,2} &=& (-4 + 15 c_2^2 - 4 c_2^3 - 8 c_3  + 3 c_4) /2 \nn \\ 
S_{3,4} &=& -18 + 44 c_2  - 16 c_2^2 - 3 c_2^3 - 7 c_2^4 + 5 c_2^5 /4 \nn \\
     &-& 22 c_3  + 12 c_2  c_3  + 10 c_2^2 c_3  + c_2^3 c_3  - 4 c_3^2 \nn \\
     &-& 7 c_2  c_3^2 /4 + 5 c_4  + 9 c_2  c_4 /8 - 9 c_2^2 c_4 /4   \nn \\
     &+& c_3  c_4  - 2 c_5  + 3 c_2  c_5 /4 + 3 c_6 /8  \nn \\
S_{4,0} &=& 12 c_2 ^2 + 4 c_3   \nn \\
S_{4,2} &=&  6 - 36 c_2  + 15 c_2^2 + 60 c_2^3 - 15 c_2^4 - 4 c_3  \nn \\
    &-& 20 c_2  c_3  - 6 c_2^2 c_3  - 5 c_4  + 18 c_2  c_4  + 2 c_5  \nn \\
S_{4,4} &=& (180 - 984 c_2  + 1488 c_2^2 - 534 c_2^3 + 99 c_2^4 \nn \\
        &-& 180 c_2^5 + 27 c_2^6 + 260 c_3  - 816 c_2  c_3  \nn \\
        &+& 232 c_2^2 c_3  + 144 c_2^3 c_3  + 30 c_2^4 c_3  + 76 c_3^2 \nn \\
        &-& 48 c_2  c_3^2 - 30 c_2^2 c_3^2 - 4 c_3^3 - 86 c_4  \nn \\ 
        &+& 163 c_2  c_4  + 129 c_2^2 c_4  - 66 c_2^3 c_4  - 72 c_3 c_4 \nn \\ 
        &+& 8 c_2  c_3  c_4  + 16 c_4^2 + 8 c_5  - 40 c_2  c_5   \nn \\
      &+& 12 c_2^2 c_5  + 7 c_3  c_5  - 5 c_6  + 12 c_2  c_6  + c_7)/2 \nn \\
S_{5,0} &=& 60 c_2 ^3 + 60 c_2  c_3  + 5 c_4 \nn \\  
S_{5,2} &=& -24 + 180 c_2  - 510 c_2^2 + 240 c_2^3 + 450 c_2^4 \nn \\
      &-& 108 c_2^5 - 80 c_3  + 90 c_2^2 c_3  - 120 c_2^3 c_3  \nn \\ 
      &-& 90 c_3^2 - 10 c_4  - 105 c_2  c_4 /2 + 170 c_2^2 c_4  \nn \\ 
      &+& 50 c_3  c_4  - 6 c_5  + 40 c_2  c_5  + 5 c_6/2 \nn \\
S_{6,0} &=& 6 (60 c_2^4 + 120 c_2^2 c_3  + 15 c_3^2 + 
              20 c_2  c_4  + c_5) \nn \\  
S_{6,2} &=& 120 - 1080 c_2  + 3870 c_2^2 - 6930 c_2^3 + 3060 c_2^4 \nn \\ 
      &+& 3600 c_2^5 - 840 c_2^6 + 480 c_3  - 3240 c_2  c_3   \nn \\
      &+& 1500 c_2^2 c_3  + 3840 c_2^3 c_3  - 1800 c_2^4 c_3  \nn \\
      &-& 110 c_3^2 - 1620 c_2  c_3^2 - 225 c_2^2 c_3^2 + 30 c_3^3 \nn \\ 
      &-& 150 c_4  - 135 c_2  c_4  - 60 c_2^2 c_4  + 1500 c_2^3 c_4  \nn \\ 
      &-& 420 c_3  c_4  + 1500 c_2  c_3  c_4  + 105 c_4^2 - 18 c_5  \nn \\ 
      &-& 108 c_2  c_5  + 585 c_2^2 c_5  + 135 c_3  c_5  - 7 c_6   \nn \\
      &+& 75 c_2  c_6  + 3 c_7  .
\label{scgic}
\eea
Note that all the values of $c_n$ can be obtained (or rewritten) in terms
of the tree-level alone, $S_{J,0}$, indicating that the knowledge
of the tree-level is enough to fully 
specify the underlying local-density transformation,
and, therefore, to generate all higher order corrections.

Furthermore, we can make use of the general formula for the top-hat filtering 
derived in \S\ref{sec:smooth}, and generalize the above given expressions
by replacing the unsmoothed $c_k$ coefficients by their smoothed counterparts
$\overline{c_k}$ in the following way:
\bea
\overline{c_2} &=& c_2 +{\gamma_1 \over 3}  \nn \\
\overline{c_3} &=& c_3 - {\gamma_1\over 2} + {3 \over 2}\,c_2\,\gamma_1 
       + {\gamma_1^2 \over 4} + {\gamma_2 \over 6}  \nn \\
\overline{c_4} &=& c_4 + {4\over 3}\,\gamma_1 - 4\,c_2\,\gamma_1
+ 2\,c_2^2\,\gamma_1 + {8\over 3}\,c_3\,\gamma_1 \nn \\
&-& {4 \over 3}\,\gamma_1^2 + {8 \over 3}\,c_2\,\gamma_1^2 
+{8\over 27} \,\gamma_1^3 - {2\over 3} \,\gamma_2 
+{4\over 3}\,c_2\,\gamma_2 \nn \\ 
&+& {4\over 9} \,\gamma_1 \gamma_2 + {2\over 27} \,\gamma_3 ,
\label{tophatgm}
\eea
and so on, where:
\bea
\gamma_p = \gamma_p(R) = {d^p\, \log \sigmahat_l^2  \over d\, \log^p R} \nn .
\label{eq:gamma}
\eea
These expressions are valid for arbitrary dynamics (to be specified 
through the unsmoothed coefficients) and a generic initial power spectrum. 
They can be used for the particular cases of interest.
For the SC dynamics ($c_k = \nu_k$ as given by Eq.[\ref{nusc}]) we find 
for the first coefficients of the cumulants,
\bea
s_{2,4} &=& {1909 \over 1323} + {143\over 126}\,\gamma_1 + 
      {11\over 36}\,\gamma_1^2 + {\gamma_2 \over 6} \nn \\
S_{3,0} &=& {34\over 7} +\gamma_1 \nn \\  
S_{3,2} &=& {1026488\over 101871} + {12862\over 1323}\,\gamma_1 
          + {407\over 126}\,\gamma_1^2 + {10\over 27}\,\gamma_1^3 \nn \\
        &+& {11 \over 7}\,\gamma_2 + {2\over 3}\,\gamma_1 \gamma_2 
         + {\gamma_3 \over 9} \nn \\
S_{4,0} &=& {60712\over 1323} + {62\over 3}\,\gamma_1 + 
{7 \over 3}\,\gamma_1^2 + {2 \over 3}\,\gamma_2 \nn \\
S_{4,2} &=& {22336534498\over 83432349} + {42649448 \over 130977}\,\gamma_1 + 
        {3571621\over 23814}\,\gamma_1^2 \nn \\
&+& {35047 \over 1134}\,\gamma_1^3 
+ {1549\over 648}\,\gamma_1^4 + {575777\over 11907}\,\gamma_2 + 
{5981\over 189}\,\gamma_1\,\gamma_2 \nn \\
&+& {263\over 54}\gamma_1^2 \gamma_2 + 
        {25\over 54}\gamma_2^2 + {2084\over 567}\,\gamma_3 +
{86\over 81}\gamma_1 \gamma_3+{5\over 81}\gamma_4 \nn \\
\label{nlscgp}
\eea
This way, the results for the SC and SSZA dynamics described in the text 
(see \S\ref{sec:comparison} and Appendix \ref{app:za} respectively) are given as particular
cases of this {\em local-density} transformation in Lagrangian space whenever
we replace the $c_k$ coefficients by those associated to the relevant
dynamics: $c_k= \nu_k$ or $c_k= \overline{\nu_k}$ for the unsmoothed  or
smoothed fields respectively (the previous being a particular case 
of the latter). The same expressions hold for the velocity divergence fields, 
replacing $c_k = \mu_k$ or $c_k = \overline{\mu_k}$ for the unsmoothed and
smoothed fields respectively. These $\mu_k$ coefficients are related to those
for the density field through the equation of continuity 
(see Paper III for details).

\section[Cumulants in Euler Space]
{Cumulants from the Local-Density Transformation of 
Gaussian Initial Conditions in Euler Space}
\label{app:euler}

In this section we derive the cumulants from the local-density transformation
of GIC in Euler space to quantify the departures form the Lagrangian 
formulation and give values for the SC approximation. 
The formulae given below reproduce
those given in Fry \& Gazta\~naga (1993). There, 
they were presented  
as a bias transformation between the luminous and 
the underlying matter fluctuations.
They can also be obtained by simply
replacing $F_n = c_n/n!$ in the expressions for the loop
corrections (in Euler space) given in SF96a.

The first perturbative contributions to the cumulants in Euler space are
the following,
\bea
s_{2,4} &=&  {c_2^2 \over 2} + c_3 \nn \\   
\xi_{3,6} &=& c_2^3 + 6 c_2 c_3 + {3\over 2} c_4 \nn \\
S_{3,0} &=& 3 c_2  \nn \\ 
S_{3,2} &=& -2 c_2^3 + {3\over 2} c_4  \nn \\
S_{4,0} &=& 12 c_2 ^2 + 4 c_3   \nn \\
S_{4,2} &=& -15 c_2^4 - 6 c_2^2 c_3 + 18 c_2 c_4 + 2 c_5  . 
\eea
where we have also computed the third order cumulant, $\xi_3$ whose
one-loop contribution is denoted by $\xi_{3,6}$ following the notation
introduced in \S\ref{sec:comparison}. It is easy to see
that this contribution is given by $\xi_{3,6} = S_{3,2}+2 s_{2,4} S_{3,0}$.

Replacing in these expressions the values for the SC dynamics with 
a top-hat filter and a power-law spectrum, \ie introducing the
smoothed coefficients given by Eqs.[\ref{nusm}],[\ref{nusc}], 
we get estimates for the cumulants as summarized in Table \ref{euler}.
For a comparison of both the Lagrangian and Eulerian predictions
within the SC model with respect to the exact PT values we also
display the results by SF96a, SF96b and S97 obtained using the 
diagrammatic representation for PT.  

Notice that unlike the case for the SC estimates from Lagrangian space,
the Euler estimates just give the right order of magnitude for the cumulants
when compared against exact analytic calculations (see Eqs.[\ref{exactpt}],
[\ref{sfn2}]).
The observed trend is that (non-reduced) cumulants in Euler space  
typically 
overestimate those in Lagrangian space as the normalization factor
(see Eq.[\ref{normalag}]) between the two spaces introduces terms that
give a negative {\em non-local} net contribution in terms of the variance
(see \S\ref{sec:comp_ept} for a discussion).
  
Since this overestimation effect in the
higher-order cumulants scales less steeply than hierarchically with respect to 
that of the variance, $\Delta \xi_J \simlt (\Delta \xi_{2})^{J-1}$, 
the effect in the $S_J$ ratios is thus dominated by 
that in the variance. This results in a net underestimate of the $S_J$ ratios.
Notice that the relative underestimation becomes larger and larger as 
smoothing effects increase.

\begin{table}

\begin{center}

\begin{tabular}{|c||c|c|c|c|}
\hline \hline
SC & Unsmoothed & \multicolumn{3}{c|}{Smoothed} \\ 
\hline \hline
& $\gamma=0$ & $\gamma=-1$ & $\gamma=-2$ & $\gamma=-3$ \\ \hline
& $n=-3$ & $n=-2$ & $n=-1$ & $n=0$  \\ 
\hline \hline
$s^E_{2,4}$   & 4.92 & 2.76 & 1.20 &  0.26 \\ 
\hline 
$s^L_{2,4}$   & 1.44 & 0.61 & 0.40 &  0.79 \\ 
\hline 
$s^{PT}_{2,4}$   & 1.82 & 0.88 & ? &  ? \\ 
\hline\hline
$\xi^E_{3,6}$   & 54.64 & 21.80 & 5.68 &  0.38 \\ 
\hline 
$\xi^L_{3,6}$   & 24.09 & 7.95 & 2.85 &  2.91 \\ 
\hline 
$\xi^{PT}_{3,6}$   & 27.49 & 9.97 & ? &  ? \\ 
\hline\hline
$S^E_{3,2}$ & 6.85 & 0.54 & -1.21 & -0.60 \\ 
\hline 
$S^L_{3,2}$ & 10.08 & 3.21 & 0.59 & -0.02 \\ 
\hline 
$S^{PT}_{3,2}$ & 9.80 & 3.18 & ? & ? \\ 
\hline\hline
$S^E_{4,2}$ & 208.45 & 24.83 & -10.71 & -3.09 \\ 
\hline
$S^L_{4,2}$ & 267.72 & 63.56 & 7.39 & -0.16 \\ 
\hline
$S^{PT}_{4,2}$ & ? & ? & ? & ? \\
\hline \hline
\label{euler_tab}

\end{tabular}

\caption[junk]{Comparison between the cumulants in the SC model
in Euler space ($^E$) with those in Lagrangian space ($^L$) and in exact
PT ($^{PT}$).}

\label{euler}
\end{center}

\end{table}

\section{Top-Hat Smoothing in Fourier Space}
\label{app:nu2}

 To illustrate the fact that the SC model gives the correct smoothed 
tree-level amplitudes for Gaussian initial conditions for 
a top-hat window, we shall derive 
$\overline{\nu_2}$ explicitly by
imposing the {\em spherically symmetric} approximation in Fourier space 
and show that the skewness
$S_3 = 3 \overline{\nu_2}$, exactly reproduces the leading order
exact perturbative result, Eq.[\ref{nusm}]: 
\eg $\overline{\nu_2}=\nu2 +\gamma/3$
We perform the calculation in Euler space but the result does not differ
from that in Lagrangian space at tree-level for Gaussian initial 
conditions, as discussed in \S\ref{sec:sc}.
Differences are only expected to appear in
higher perturbative orders ($\sigma$-corrections).

To see this, we recall the properties derived by Bernardeau 1994b 
(B94b henceforth)
for a top-hat window function (spherical window) defined as,
\bea
W_{TH} (\x) = 1 \quad if \quad |x| \leq R_0 \nn ,
\eea
and $0$ otherwise, for a scale $R_0$, so that the smoothed fields are
obtained through the integral,
\bea
\overline{\delta} \equiv \delta (R_0) = \int d^3 \x\, W_{TH}(\x)\delta(\x) \nn ,
\eea
We turn to Fourier space for convenience, where the smoothed fields 
are expressed as,
\bea
\delta (R_0) = \int {d^3 \k \over (2\pi)^{3/2}}\,W_{TH}(\k R_0)\delta(k) \nn ,
\eea
where $W_{TH}(\k R_0)$ is the Fourier transform of $W_{TH}(\x)$.
In particular, for the second-order in the perturbative series, we have for the
smoothed field 
\bea
\delta_2 (R_0) = \int {d^3 \k_1 \over (2\pi)^{3/2}}\,
{d^3 \k_2 \over (2\pi)^{3/2}}\,\delta_{k_1}\,\delta_{k_2}\,
W(|\k_1+\k_2|\,R_0)\, \nn \\ 
\times \left[D_1^2\,\left(P_{1,2}-{3\over 2}Q_{1,2}\right)
\,+\,{3\over 4} D_2\,Q_{1,2}\right] \nn ,
\eea
where,
\bea
P_{1,2} = 1\,+\,{\k_1\cdot \k_2 \over k_1^2}, \quad
Q_{1,2} = 1\,+\,{(\k_1\cdot \k_2)^2 \over k_1^2\,k_2^2} \nn  ,
\eea
and $D_i$ are the time-dependent growth factors for the $i$-th perturbative
order to be solved with the SC equations of motion.
Now, decomposing the integrals into their angular and radial part
and translating the property of spherical symmetry into  
Fourier space language,
\beq
\delta_{k_i} = \delta_{|k_i|}  , 
\eeq
we can apply some properties for the top-hat window function 
(see B94b, Eqs.[A5] and [A6]) and get,
\bea
\delta_2(R_0) = \left({D_2\over 2\,D_1^2}\,+\,{1\over 6}{d\,
\log[\delta_l(R_0)]^2 \over d\,\log R_0}\right)\,[\delta_l(R_0)]^2 \nn , 
\eea
since,
\bea
\delta_l (R_0) = \int {d^3 \k \over (2\pi)^{3/2}}\,W_{TH}(\k R_0)
\delta_{k}\,D_1(t) \nn  .
\eea
We further use the fact that in general 
the smoothed linear density 
contrast at a point $x$, can be factorized in its scale dependent part 
$\sigma (R_0)$ and its normalized
linear field $\varepsilon(x)$,
$\delta_l (R_0) \equiv \delta_l(R_0,x) = \sigma_l (R_0)\,\eta(x)$,
and
we finally get:
\bea
\delta_2(R_0) 
= {1\over 2}\,\left({34\over 21}\,+{1\over{3}} \,{d\,
\log[\sigma_l(R_0)]^2 \over d\,\log R_0}\right)\,[\delta_l(R_0)]^2. 
\eea
If we define,$\delta_2(R_0) 
\equiv (\overline{\nu_2}/2)\,~[\delta_l(R_0)]^2$,
we find
\beq
\overline{\nu_2}= {34\over 21}\,+{1\over{3}} \,{d\,
\log[\sigma_l(R_0)]^2 \over d\,\log R_0} = \nu_2 +{\gamma\over 3}
\eeq
which exactly reproduces Eq.[\ref{nusm}].
This result can be extended to higher orders following the
properties of the top-hat window presented in B94b.

\section{The Spherically Symmetric Zel'dovich Approximation}
\label{app:za}

A simple accurate description of the non-linear dynamics of the
fluid elements before shell crossing (single streaming regime) is provided 
by the so-called Zel'dovich 
approximation (hereafter ZA, see Zel'dovich 1970). 
According to this approximation, particle 
positions in comoving coordinates are assumed to follow
straight lines. Despite being exact only at linear order in Lagrangian PT,  the
latter solution has been successfully tested as an accurate approximation
for the description of the dynamics in the non-linear regime (see
\eg Croft \& Gazta\~{n}aga 1998).
 Within the ZA the fluid equations can then be 
easily integrated, and yield
\beq
1+\delta = \prod_{i=1}^{N} {1\over{(1- D(t)\,\lambda_i)}} ,
\label{zeltrans} 
\eeq
where $N$ is the number of spatial dimensions and
$\lambda_i = -{\partial \psi_i /{\partial q_i}}$,
which is in general a {\em non-local} relation between the 
displacement field and
the evolved density fluctuation.

At this point we introduce the {\em spherically symmetric} assumption 
for the ZA dynamics (SSZA) as we did before with the exact dynamics.
Thus we set $\lambda_i = \lambda$,
\ie all directional derivatives are equal due to the spherical symmetry 
(no tidal forces) and rewrite,
$\delta_l = D(t)\, \sum_{i=1}^{N} \lambda_i = 3\,D(t)\,\lambda$,
as follows from linearising Eq.[\ref{zeltrans}].
The assumption of spherical symmetry renders the relation 
Eq.[\ref{zeltrans}] as a local
transformation that may be expanded in power series 
to get the full perturbative
series for the density contrast. In particular, for the
3D case, the normalized SSZA transformation has the form,
\beq
\delta \sim (1-\delta_l/3)^{-3} ,
\label{za3d}
\eeq 
to be normalized according to Eq.[\ref{normalag}].
Notice that the latter transformation may be straightforwardly extended to any
spatial dimensions. As seen in Fosalba \etal (1998), the exponential 
transformation
that leads to the lognormal hierarchical amplitudes, $S_J = J^{J-2}$, are
only recovered when one takes the limit of infinite spatial dimensions
(see also Bernardeau and Kofman 1995), which
does not make much physical sense.

Expanding Eq.[\ref{za3d}] in power series we can determine the
unsmoothed coefficients $\nu_k$ of the local non-linear transformation 
for the SSZA, 
\bea
&&\nu_2 = {4\over 3} \sim  1.33; \quad 
\nu_3 = {20\over 9} \sim  2.22 \nn \\ 
&&\nu_4 = {40\over 9} \sim 4.44; \quad
\nu_5 = {280\over 27} \sim 10.37 ,
\label{nuzasc}
\eea
and so forth. 
Introducing them in the smoothed transformation according to 
Eq.[\ref{nusm}] for a top-hat window, we can now find, using
Eq.[\ref{tophatgm}] with $c_k=\overline{\nu_k}$,
the following results for the smoothed density field for GIC,
and a power-law power spectrum,
\bea
s_{2,4} &=& {7\over 9} + {11\over 18}\,\gamma + 
{11 \over 36}\,\gamma^2   \nn \\ 
s_{2,6} &=& {185\over 243} + {535\over 324}\,\gamma + {1043 \over 648}
\,\gamma^2 + 
{127 \over 162}\,\gamma^3  
+ {127 \over 648}\,\gamma^4 \nn \\
S_{3,0} &\equiv& S_3^{(0)} = 4 + \gamma \nn \\
S_{3,2} &=&
{118 \over 27} + {16\over 3}\,\gamma + 
{41 \over 18}\,\gamma^2 + 
{10 \over 27}\,\gamma^3  \nn \\ 
S_{4,0} &\equiv& S_4^{(0)} = {272\over 9} + {50 \over 3}\gamma + 
{7 \over 3}\gamma^2  \nn \\
S_{4,2} &=& {2506\over 27} + {11512 \over 81}\gamma  
+  {13523 \over 162}\gamma^2 + 
{3679 \over 162}\gamma^3 + {1549 \over 648}\gamma^4
\nn \\   
\eea
The results for the variance and $S_J$ for the SSZA for particular 
values of the spectral index are given in Table \ref{nlza}.

For the {\em unsmoothed} fields ($\gamma=0$), analytic results including the 
two corrective terms beyond the tree-level, were derived by 
SF96a in the context of the diagrammatic approach. 
For the variance, skewness, and the
kurtosis (the latter only up to the first $\sigma$-correction) they give,
\bea
\sigma^2 \approx \sigma_l^2\,+\,1.27\,\sigma_l^4\,+\,2.02\,\sigma_l^6\,+\,
\Or \left(\sigma_l^8 \right) \nn \\ 
S_3 \approx 4\,+\,4.69\,\sigma_l^2\,+\,13.93\,\sigma_l^4\,+\,
\Or \left(\sigma_l^6 \right) \nn \\
S_4 \approx 30.22\,+\,98.51\,\sigma_l^2\,+\,
\Or \left(\sigma_l^4 \right)  .
\eea
These results are to be compared to our results from the
SC model in the perturbative regime (truncated at the 
same order),
\bea
\sigma^2 \approx \sigma_l^2\,+\,0.78\,\sigma_l^4\,+\,0.76\,\sigma_l^6\,+\,
\,\Or \left(\sigma_l^8 \right) \nn \\
S_3 \approx 4\,+\,4.37\,\sigma_l^2\,+\,\,9.41\,\sigma_l^4\,+\,
\Or \left(\sigma_l^6 \right) \nn \\
S_4 \approx 30.22\,+\,92.81\,\sigma_l^2\,+\,331.0\,\sigma_l^4\,+\,
\,\Or \left(\sigma_l^6 \right) ,
\eea
which differ in $7\,\%$ in the corrective term for $S_3$, 
$6\,\%$ in the corrective term for $S_4$, and up to $40\,\%$ in
the one-loop term for the variance. The two-loop contributions
for $S_3$ differ in about $30\,\%$ and just give the right order of 
magnitude for the variance.  The two-loop contribution to $S_4$ 
($\sim 331\,\sigma_l^2$) must be taken just as an estimate of the actual
value for this coefficient in PT, 
since there are no analytic results available to 
compare to. All the above quoted differences must be attributed to the
tidal effects as argued before what gives further support to the view that
the hierarchical amplitudes are essentially of a {\em shearless} nature unlike
the (unreduced) one-point cumulants. 

For the smoothed density field in the ZA to PT there is only one
result available concerning the skewness for $n=-2$ (see S97),
\beq
S_3 \approx 3 + 0.82\,\sigma_l^2 + \Or (\sigma_l^4),
\eeq
while within the SSZA dynamics, we obtain,
\beq
S_3 \approx 3 + 0.94\,\sigma_l^2 + \Or (\sigma_l^4),
\eeq 
which means a $15 \%$ negative contribution from the shear
for the first corrective term for that
particular value of the spectral index.

\begin{table}

\begin{center}

\begin{tabular}{|c||c|c|c|c|}
\hline \hline
SSZA & Unsmoothed & \multicolumn{3}{c|}{Smoothed} \\ 
\hline \hline 
& $\gamma=0$ & $\gamma=-1$ & $\gamma=-2$ & $\gamma=-3$ \\ \hline
& $n=-3$ & $n=-2$ & $n=-1$ & $n=0$ \\ 
\hline \hline
$s_{2,4}$   & 0.78 & 0.47 & 0.78 & 1.69 \\ 
\hline \hline
$S_{3,0}$ & 4 & 3 & 2 & 1 \\ 
\hline
$S_{3,2}$ & 4.37 & 0.94 & $-0.15$ & -1.13 \\ 
\hline \hline
$S_{4,0}$ & 30.22 & 15.89 & 6.22 & 1.22 \\ 
\hline
$S_{4,2}$ & 92.81 & 13.85 & $-0.96$ & -1.82 \\ 
\hline \hline 
\end{tabular}

\caption[junk]{Values for the higher-order perturbative contributions 
for the SSZA for the unsmoothed ($n=-3$) and smoothed ($n=-2,-1,0$) 
density fields, for a top-hat window and a power-law power spectrum.}
\label{nlza}
\end{center}

\end{table}

\begin{figure}[t]
\centering
\centerline{\epsfysize=8.truecm 
\epsfbox{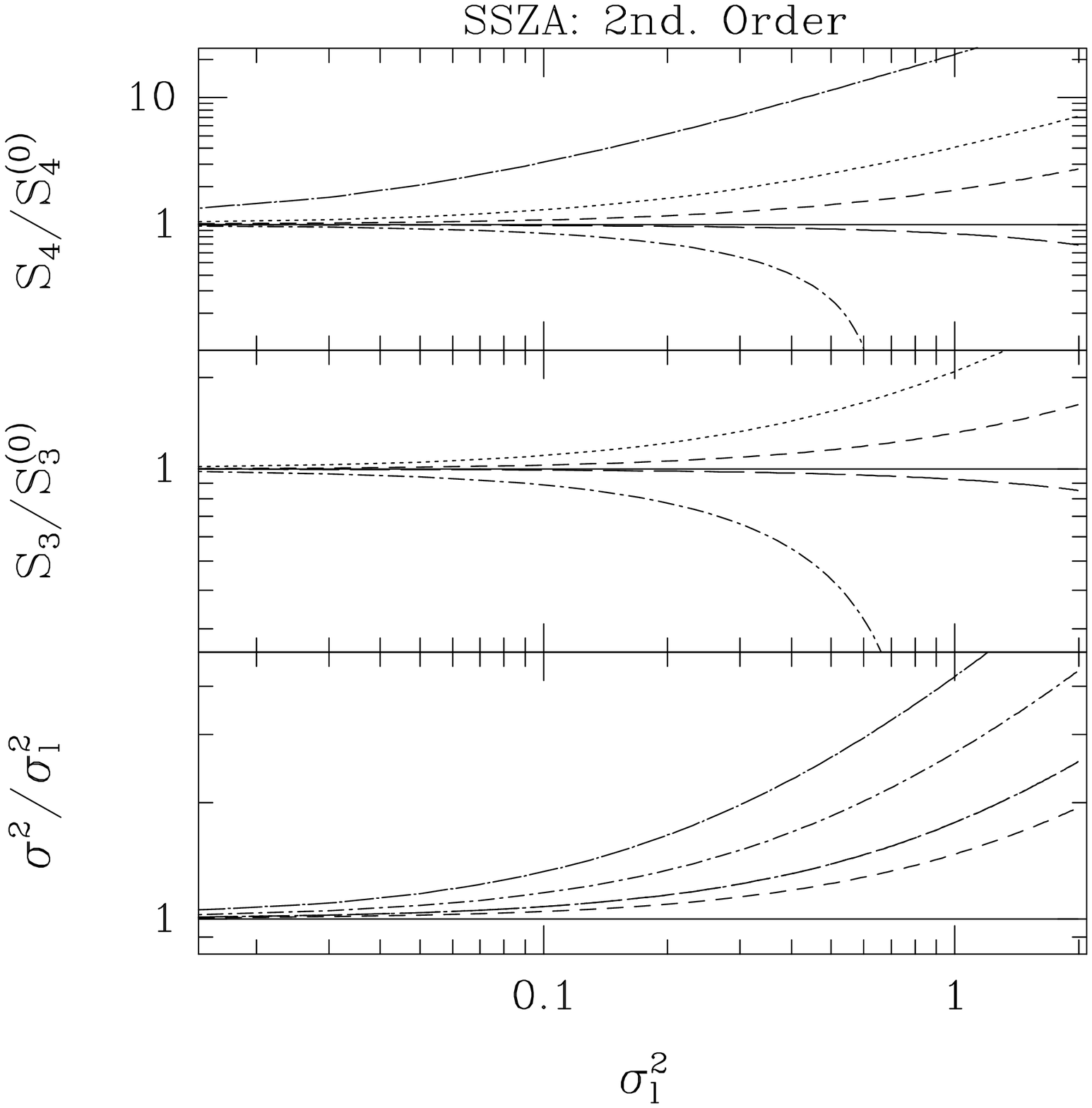}}
\caption[junk]{Same as Fig \ref{scvpt1}, for the SSZA. 
In this case, the variance has a minimum corrective term for 
$n\approx -1$, and the hierarchical amplitudes show a vanishing first 
corrective 
term for $n \approx -1.2$. Notice that the $n=1$ line is not plotted for the
skewness since
for that particular value the tree-level exactly vanishes.} 
\label{scvptza1}
\end{figure}


In Fig \ref{scvptza1} we display the deviations from the
tree-level values for the variance, skewness and kurtosis for the 
SSZA up to 2nd perturbative contribution (one-loop).
In line with the arguments pointed out for the spherically symmetric 
approximation to the exact PT (the SC model), here the smoothing effects
tend to diminish the non-linear corrections as well. We have checked
that the 3rd order contribution follows qualitatively the same behavior
although non-linearities are typically larger. 
On the other
hand, the lack of a critical index (vanishing non-linearities) for the
variance gives further support to our claim that it is due to the local 
nature of the SC picture and thus, to the loss of the previrialization effect.

\end{document}